\documentclass[preprintnumbers,article,amsmath,amssymb,floatfix,10pt,prd,superscriptaddress,nofootinbib]{revtex4}

\usepackage{doi}
\usepackage{hyperref}
\hypersetup{
  colorlinks=true,        
  linkcolor=magenta,         
  citecolor=red,         
}

\usepackage{graphicx}
\usepackage{dcolumn}
\usepackage{bm}
\usepackage{color}
\usepackage{enumitem}
\usepackage{amsmath}
\usepackage{amssymb}

\newcommand{\beq}{\begin{equation}}
\newcommand{\eeq}{\end{equation}}
\newcommand{\bea}{\begin{eqnarray}}
\newcommand{\eea}{\end{eqnarray}}

\usepackage{bbm}
\usepackage{amsfonts}
\usepackage{mathrsfs}
\usepackage{latexsym}
\usepackage{epsfig}
\usepackage{epstopdf}
\usepackage{epstopdf}
\usepackage{graphicx}
\usepackage{amssymb}
\usepackage{amsmath}
\usepackage{dcolumn}
\usepackage{bm}
\usepackage{color}
\usepackage{comment}
\usepackage{xcolor}
\usepackage{xcolor}
\usepackage{orcidlink}

\begin{document}
\title{Testing $(q)$-Deformed Dunkl-Fokker-Planck Equation Algebra with\\ Supersymmetry (SUSY) and Foldy-Wouthuysen (FW) Measurement}

\author{Abdelmalek Bouzenada\orcidlink{0000-0002-3363-980X}}
\email{abdelmalekbouzenada@gmail.com}
\affiliation{Laboratory of Theoretical and Applied Physics, Echahid Cheikh Larbi Tebessi University, 12001, Algeria}
\affiliation{Research Center of Astrophysics and Cosmology, Khazar University, Baku AZ1096, Azerbaijan}

\begin{abstract}
In this study, a relativistic formulation of the $(q)$-deformed Dunkl-Fokker-Planck equation in $(1+1)$-dimensions is constructed within the reflection-deformed quantum framework. In this case, the formalism includes $(q)$-deformed Dunkl operators and reflection symmetry to build a generalized dynamical structure for a relativistic quantum systems framework. Moreover, the corresponding $(q)$-Wigner-Dunkl supersymmetric configuration is established via the construction of deformed ladder operators and supersymmetric algebraic relations, yielding a consistent spectral representation of the model within the algebraic framework. The analysis extends to the harmonic oscillator with centrifugal interaction, where exact algebraic solutions, similarity reduction techniques, and closed energy spectra are obtained analytically in detail. The role of the deformation parameter and reflection operator on spectral properties and wavefunction structure is examined systematically in detail. A generalized Foldy-Wouthuysen (FW) transformation is introduced within the deformed Dunkl framework to achieve relativistic decoupling of positive- and negative-energy sectors within the present theoretical formulation. In this case, this approach yields an effective reduced Hamiltonian, including higher-order relativistic and deformation-induced terms. Also, the associated Dunkl-Fokker-Planck dynamics generated through high-order FW reduction are examined in detail for reflection-deformed relativistic quantum systems. In this context, results obtained here yield a unified algebraic and relativistic description of $(q)$-deformed Dunkl structures and construct a consistent framework for investigating supersymmetric and relativistic properties in reflection-symmetric quantum models in general.
\end{abstract}

\maketitle

\date{\today}


\section{Introduction}

Quantum mechanics (QM) \cite{a1} and general relativity (GR) \cite{a2} characterize physical phenomena across different scales. Moreover, QM governs microscopic systems exhibiting probabilistic dynamics and non-classical effects such as superposition and entanglement \cite{a3, a4, a5}, and it is widely used in condensed matter physics, semiconductor technology, and quantum information processing. In this case, GR, formulated by Einstein, describes gravitation via spacetime curvature produced by energy and mass, accounting for phenomena associated with black holes and cosmological evolution \cite{a6}. However, these two theories become incompatible under extreme conditions near the Planck scale \cite{a7}, motivating the development of quantum gravity frameworks such as string theory and loop quantum gravity \cite{a8}. Also, in relativistic quantum field theory in curved spacetime \cite{BZ22, BZ23, BZ24, BZ25, BZ26, BZ27, BZ28, BZ29, BZ30, BZ31, BZ32}, particle dynamics depend on intrinsic spin and background geometry. Spin-0 scalar and spin-1/2 fermionic fields exhibit distinct couplings to curvature, yielding different dynamical equations and energy spectra compared with flat spacetime cases \cite{a9, a10}.

Supersymmetry (SUSY) is defined as a graded symmetry relating internal symmetries and spacetime symmetries within a non-trivial theoretical framework. Although supersymmetric particles have not been observed experimentally so far, this situation is interpreted in terms of SUSY breaking occurring at energy scales above the current particle accelerator reach limits. In this case, SUSY may still contribute at high-energy early-universe regimes, where restoration of SUSY can produce observable signatures in primordial cosmological perturbations \cite{SSy2} within inflationary cosmology frameworks during primordial evolution models. In local SUSY, supergravity merges general relativity into a unified formulation and establishes a theoretical basis for cosmological analysis in FLRW geometries within a homogeneous isotropic spacetime sectors framework. Also, the supersymmetric structure of gravity contributes directly to quantum cosmology since quantization introduces additional supersymmetric constraints that can restrict admissible quantum states and, in certain cases, produce unique wave functions interpreted as intrinsic boundary conditions of the theory framework constraints structure approach. Multiple formulations of supersymmetric cosmology are developed in the literature context frameworks. One relevant approach relies on the reduction of degrees of freedom in $N=1$ four-dimensional supergravity while maintaining spatial homogeneity and isotropy, yielding effective cosmological models with extended local SUSY within a reduced dynamics framework. SUSY quantization has been studied in anisotropic scalar-tensor cosmologies where supersymmetric constraints modify the quantum structure of cosmological dynamics at the effective level analysis framework. In addition, supersymmetric FRW cosmological models coupled to scalar and tachyonic fields are analyzed in relation to inflationary scenarios, quantum wave functions, and dark energy dynamics framework studies. Higher-derivative supersymmetric cosmological models have been proposed to study corrections from generalized gravitational actions and their effects on the quantum evolution of the universe framework through theoretical analysis. In this context, advanced developments in supersymmetric quantum cosmology, covering wave function structure, boundary conditions, and semiclassical limits, are discussed in Refs. \cite{SSy8, SSy9, SSy10} comprehensive theoretical review framework.

Several approaches are constructed to extend Newtonian and quantum mechanics \cite{BZ1, BZ2, BZ3, BZ4, BZ5, BZ6, BZ7, BZ8, BZ9, BZ10} through the introduction of deformed differential operators replacing the standard time derivative operator structure. In these formulations, ordinary derivative $(d/dt)$ is substituted with a generalized operator $(D_t^{A})$ dependent on deformation parameter $(A)$, yielding modified definitions of velocity and dynamical evolution. Also, such constructions yield effective descriptions for complex physical systems, especially when nonlocal effects or additional internal structures occur, while the classical formulation is recovered in the appropriate limit of the deformation parameter within the theoretical framework space \cite{DO1, DO2, DO3, DO4, DO5}. A distinct and important generalization relies on fractional calculus, where differentiation extends to non-integer orders in mathematical physics. Fractional derivatives are widely employed in the formulation of generalized mechanical systems characterized by memory effects, anomalous transport, and viscoelastic behavior processes considered \cite{DO5, DO6, DO7, DO8, DO9, DO10}. In this framework, standard kinematic quantities are modified accordingly, and the resulting fractional mechanics provides a consistent extension of classical dynamics for systems with nonlocal temporal behavior in theoretical modeling. The classical Newtonian formulation is recovered when the fractional order approaches unity, ensuring consistency with the standard mechanics limit condition. Fractional approaches have been extensively applied in the description of effective physical models involving dissipation and complex media systems studies \cite{DO11, DO12, DO13, DO14}. Among various generalized derivative operators, the Dunkl operator plays a distinct role due to its intrinsic connection with the reflection symmetry and the framework of discrete invariance properties. Originally introduced in the study of differential-difference operators associated with reflection groups, the Dunkl derivative combines the ordinary derivative with a reflection operator weighted by a deformation parameter in algebraic systems \cite{DO15}. This structure yields a nonlocal operator preserving parity-type symmetry and modifying underlying algebraic properties of the system framework characteristics. Also, the mathematical foundations of Dunkl theory were further developed in the context of orthogonal polynomials and harmonic analysis, establishing a rich framework for applications in mathematical physics research \cite{DO16}. The Dunkl operator has also been linked to generalized formulations of quantum mechanics through modified commutation structures and oscillator models \cite{DO17}. Its algebraic structure has been extensively investigated in relation to integrable systems, affine Hecke algebras, and representation theory \cite{DO18, DO19}. These developments show that Dunkl formalism provides a natural extension of conventional differential operators for systems with reflection symmetries and a framework of hidden algebraic structures. Moreover, from a physical perspective, Dunkl derivatives have been widely applied in the study of exactly solvable quantum systems and interacting model theory analysis. In particular, they appear in quantum many-body systems associated with root structures and integrable interactions, where they modify the kinetic term and introduce an effective reflection-dependent interaction mechanism \cite{DO20, DO21, DO22}. In this case, this formalism has been successfully used in the construction of Dunkl oscillator models in one, two, and three dimensions, leading to modified spectra and nontrivial symmetry algebra results \cite{DO23, DO24, DO25}. Also, Dunkl-based approaches have been extended to Coulomb-like systems and algebraic solution methods involving coherent states and symmetry techniques \cite{DO26, DO27, DO28}. More recently, a one-dimensional quantum mechanical formulation based on the Dunkl derivative has been developed, where the momentum operator is defined directly through the Dunkl operator rather than the ordinary derivative \cite{DO29}. In this context, this framework, often referred to as Dunkl-type quantum mechanics, provides a consistent extension of standard quantum theory in the presence of reflection symmetry, leading to modified eigenvalue structures and altered dynamical behavior properties.

Quantum algebras and quantum group structures form an algebraic framework employed to describe a broad class of physical systems, particularly in situations where classical symmetry algebras are extended or deformed to include nontrivial dynamical effects in theoretical constructions \cite{BZ11, BZ12, BZ13, BZ14, BZ15, BZ16, BZ17, BZ18, BZ19, BZ20, BZ21}. Also, these structures are used in the formulation of exactly solvable models in statistical mechanics and conformal field theory, and they also provide a consistent algebraic description applicable to cosmic string spacetimes and condensed matter systems in solid-state theory within modern theoretical physics \cite{QD1, QD2, QD3, QD4}. A fundamental element in this framework is the q-deformation of the Heisenberg algebra, where the canonical commutation relations are modified through a deformation parameter $q$ that parametrizes deviations from standard quantum mechanics in operator algebra formulations. In this case, this deformation alters the algebraic relations while reproducing the conventional limit when $q \rightarrow 1$, maintaining consistency with ordinary quantum theory. From this construction, q-deformed versions of fundamental equations such as the Schrödinger equation are obtained, resulting in modified spectral properties and altered eigenvalue structures relative to the standard formulation, thereby extending the dynamical description of quantum systems. Applications of q-deformation include the q-deformed hydrogen atom and harmonic oscillator, where the deformation parameter modifies energy spectra and system dynamics. In parallel, the Weyl-Heisenberg algebra admits a formulation in the Fock-Bargmann representation, which provides an analytic description of quantum states in complex phase space. Also, this representation is applied in the analysis of squeezed states, lattice systems, and Bloch-type wave functions in periodic media, connecting algebraic methods with functional analytic techniques in quantum mechanics \cite{QD15}. The q-deformed harmonic oscillator based on the SU$_q(2)$ algebra and associated ladder operators gives a unified scheme that interpolates between bosonic and fermionic statistical behavior. Within this formulation, q-boson and q-fermion distributions arise directly from the deformed algebra, allowing continuous transitions between distinct quantum statistical regimes \cite{QD16, QD17, QD18, QD19, QD20}. In semiclassical and kinetic descriptions, these models are used to derive modified thermodynamic quantities and transport properties, where the deformation parameter controls deviations from classical statistical laws \cite{QD21}. In addition, q-calculus introduces a generalized mathematical structure for statistical mechanics in which nonextensive features originate from the deformation itself. In this setting, the parameter $q$ governs departures from Boltzmann-Gibbs statistics and introduces generalized exponential and logarithmic functions consistent with the underlying algebraic structure \cite{QD22}. In this context, a generalized thermostatistical framework has been developed, including q-deformed entropy measures for bosonic systems, which enables systematic study of deformation effects on statistical distributions and thermodynamic quantities \cite{QD23, QD24}.

A wide range of physical, chemical, and biological processes is consistently described within the framework of stochastic differential equations, where random fluctuations play a fundamental role in the dynamical evolution of the system. Within this framework, the Fokker-Planck equation occupies a central role, as it provides a deterministic evolution equation for the probability density function associated with stochastic variables and thus encodes the complete statistical behavior of the underlying random process. A representative example is Brownian motion, which has historically served as the primary physical realization for the formulation and interpretation of the Fokker-Planck equation and its generalized forms \cite{libro1,libro2,libro3,risken, junker}. The mathematical structure of the Fokker-Planck equation permits the analysis of diffusion-like phenomena, relaxation mechanisms, and noise-induced effects in complex systems, thereby serving as a versatile tool in nonequilibrium statistical physics. Over time, multiple complementary techniques have been developed to obtain explicit or approximate solutions of the Fokker-Planck equation. These methods include direct analytical integration in special cases, mapping the equation into an equivalent Schrödinger-type eigenvalue problem, and numerical implementations when closed-form solutions are not available \cite{risken}. The mapping to a Schrödinger-like equation is particularly relevant, as it enables the application of well-established tools from quantum mechanics for investigating stochastic dynamics. In this reformulated setting, powerful algebraic and symmetry-based techniques become applicable, including group-theoretical methods \cite{gt1,gt2}, supersymmetric quantum mechanics, and the principle of shape invariance \cite{junker,prl,polloto,anjos}, all of which have shown strong efficiency in constructing exact solutions for confined and exactly solvable potentials \cite{risken}. In recent developments, a broader generalized mathematical framework has emerged in quantum mechanics through the modification of the standard differential operator $(D_x \equiv \frac{\partial}{\partial x})$ into the Dunkl derivative $(D_x \equiv \frac{\partial}{\partial x} + \frac{\mu}{x}(1-R))$, where $(R)$ denotes the reflection operator acting on functions as $(Rf(x)=f(-x))$ \cite{GEN1, GEN2, GEN3, GEN4, GAZ1, GAZ2, GAZ3, SCH, SCH2}. This construction introduces an additional framework combining differential and reflection symmetries, thereby extending the mathematical description of quantum and stochastic systems. The parameter $(\mu)$ embedded in the Dunkl operator plays a central role in controlling the strength of the deformation from the standard derivative, leading to modified spectral properties and altered dynamical behavior. Such generalized operators have attracted considerable attention due to their ability to generate extended classes of exact or quasi-exactly solvable models. Moreover, the Dunkl framework has been widely applied in statistical physics, thermodynamic modeling, and related fields, where it provides a flexible tool for describing systems with nontrivial symmetry constraints and underlying discrete transformations \cite{DONGT, HAM1, HAM2, BER, QUES, QUES2, JUNK, HAM3, HAM4, HAM5, HAM6}. In particular, the introduction of the deformation parameter enables improved adaptability of theoretical predictions, providing a mechanism to fine-tune mathematical models so that they can more accurately reproduce or approximate experimental observations in complex physical environments and related stochastic systems in theoretical modeling frameworks  \cite{DONGT, HAM1, HAM2, BER, QUES, QUES2, JUNK, HAM3, HAM4, HAM5, HAM6}.

The structure of this paper is organized as follows: Section (\ref{S2}) shows the formulation of the $(q)$-deformed Dunkl-Fokker-Planck equation in $(1+1)$ dimensions, where the fundamental algebraic operators, reflection structure, and deformation scheme are introduced within a relativistic quantum framework. Section (\ref{S3}) develops the relation between the $(q)$-deformed Dunkl-Fokker-Planck equation and the $(q)$-Wigner-Dunkl supersymmetric structure, with emphasis on the associated supersymmetric algebra and reflection-deformed spectral operators. In Section (\ref{S4}), the formalism is extended to the harmonic oscillator with centrifugal interaction, where the exact algebraic structure, similarity reduction procedure, and closed spectral solutions are obtained explicitly in the deformed Dunkl representation. Section (\ref{S5}) generalizes the $(q)$-deformed Dunkl structure through the FW transformation, allowing the relativistic decoupling of positive- and negative-energy sectors and the construction of an effective reduced Hamiltonian. Also, Section (\ref{S6}) investigates the Dunkl-Fokker-Planck dynamics through high-order FW reduction in reflection-deformed relativistic quantum systems, including higher-order relativistic contributions and deformation-induced corrections. In this context, our results of the present analysis are explained and discussed in Section (\ref{S7}).

\section{The $(q)$-Deformed Dunkl-Fokker-Planck Equation in $(1+1)$-Dimensions}\label{S2}

Stochastic diffusion processes play a central role in statistical physics, quantum dynamics, transport theory, nonequilibrium thermodynamics, and mathematical physics. The standard Fokker-Planck equation governs the time evolution of the probability density for a stochastic variable under diffusion and drift mechanisms. In classical continuous geometry, the dynamics is formulated using differential operators on Euclidean space. The inclusion of quantum deformations together with reflection symmetry produces a more general structure characterized by nonlocal contributions, parity decomposition, modified spectral properties, and extended diffusion geometry. Also, the Jackson $(q)$-calculus introduces a discrete scaling symmetry that modifies the differential structure, while the Dunkl operator introduces reflection-dependent terms associated with finite Coxeter groups. In this case, the combination of both structures yields a generalized stochastic equation governed by $(q)$-deformed Dunkl operators. This framework extends the conventional diffusion model and produces exactly solvable stochastic systems with underlying algebraic and supersymmetric properties. The starting point is the one-dimensional Fokker-Planck equation written in the form \cite{Rev}:
\begin{equation}
\frac{\partial \mathcal{P}(x,t)}{\partial t}
=
\left[
-\frac{\partial}{\partial x}D^{(1)}(x)
+
\frac{\partial^2}{\partial x^2}D^{(2)}(x)
\right]
\mathcal{P}(x,t),
\label{eq1}
\end{equation}
where $\mathcal{P}(x,t)$ denotes the probability density distribution, while $D^{(1)}(x)$ and $D^{(2)}(x)$ represent the drift and diffusion coefficients. In order to construct an exactly solvable deformation scheme, the diffusion coefficient is chosen to be constant,
\begin{equation}
D^{(2)}(x)=1.
\label{eq2}
\end{equation}
The drift contribution is parameterized through a generalized $(q)$-deformed superpotential,
\begin{equation}
D^{(1)}(x)=2W_q(x),
\label{eq3}
\end{equation}
which transforms the diffusion problem into a supersymmetric spectral system. Also, the quantum deformation is introduced through the Jackson derivative,
\begin{equation}
D_qf(x)
=
\frac{f(qx)-f(x)}{(q-1)x},
\qquad q\neq1,
\label{eq4}
\end{equation}
which reduces continuously to the ordinary derivative in the undeformed limit,
\begin{equation}
\lim_{q\rightarrow1}D_qf(x)
=
\frac{df(x)}{dx}.
\label{eq5}
\end{equation}
The Jackson derivative satisfies several algebraic properties. Also, we can write,
\begin{equation}
D_qx^n=[n]_qx^{n-1},
\label{eq6}
\end{equation}
where the $(q)$-number is defined by
\begin{equation}
[n]_q=\frac{1-q^n}{1-q}.
\label{eq7}
\end{equation}
The generalized Leibniz rule becomes
\begin{equation}
D_q(fg)
=
(D_qf)g+f(qx)(D_qg),
\label{eq8}
\end{equation}
which modifies the product structure under scaling transformations. Consequently, the geometry acquires a noncommutative character with respect to dilatations.

The reflection symmetry is incorporated through the Dunkl operator. Also, the generalized $(q)$-deformed Dunkl derivative is introduced as
\begin{equation}
\mathcal{D}_{q,\mu}
=
D_q+\frac{\mu}{x}(1-R),
\label{eq9}
\end{equation}
where $\mu>-1/2$ denotes the Dunkl parameter and $R$ is the reflection operator satisfying
\begin{equation}
Rf(x)=f(-x),
\qquad
R^2=1.
\label{eq10}
\end{equation}
The operator algebra follows from the action of reflection on coordinates and derivatives,
\begin{equation}
Rx=-xR,
\qquad
RD_q=-D_qR,
\qquad
R\mathcal{D}_{q,\mu}
=
-\mathcal{D}_{q,\mu}R.
\label{eq11}
\end{equation}
The anticommutation relations separate the space into even and odd parity sectors. Therefore, the stochastic evolution depends on parity and produces distinct diffusion spectra for each sector. Also, the square of the generalized Dunkl operator determines the effective diffusion operator. Starting from
\begin{equation}
\mathcal{D}_{q,\mu}^2
=
\left(
D_q+\frac{\mu}{x}(1-R)
\right)^2,
\label{eq12}
\end{equation}
one obtains
\begin{align}
\mathcal{D}_{q,\mu}^2
&=
D_q^2
+
\frac{\mu}{x}(1-R)D_q
+
D_q\left[
\frac{\mu}{x}(1-R)
\right]
+
\frac{\mu^2}{x^2}(1-R)^2.
\label{eq13}
\end{align}

Using the relations
\begin{equation}
(1-R)^2=2(1-R),
\label{eq14}
\end{equation}
and
\begin{equation}
D_qR=-RD_q,
\label{eq15}
\end{equation}
the exact expression becomes
\begin{equation}
\mathcal{D}_{q,\mu}^2
=
D_q^2
+
\frac{2\mu}{x}D_q
+
\frac{\mu(2\mu-1)}{x^2}(1-R)
\label{eq16}
\end{equation}
Equation (\ref{eq16}) defines the $(q)$-deformed Dunkl Laplacian. The last term introduces a singular parity-dependent interaction proportional to $x^{-2}$. This contribution modifies the diffusion structure near the origin and produces different localization behavior for even and odd parity states. Also, the generalized $(q)$-deformed Dunkl-Fokker-Planck equation is therefore defined by
\begin{equation}
\frac{\partial \mathcal{P}(x,t)}{\partial t}
=
\left[
\mathcal{D}_{q,\mu}^2
-
2\mathcal{D}_{q,\mu}W_q(x)
\right]
\mathcal{P}(x,t).
\label{eq17}
\end{equation}
Substituting equation (\ref{eq16}) into equation (\ref{eq17}) gives
\begin{align}
\frac{\partial \mathcal{P}(x,t)}{\partial t} = \Bigg[ D_q^2 + \frac{2\mu}{x}D_q + \frac{\mu(2\mu-1)}{x^2}(1-R) - 2(D_qW_q(x)) - 2W_q(x)D_q - \frac{2\mu}{x}(1-R)W_q(x) \Bigg]\mathcal{P}(x,t).
\label{eq18}
\end{align}
The temporal dependence is separated through
\begin{equation}
\mathcal{P}(x,t)=e^{-\Lambda t}\Psi(x),
\label{eq19}
\end{equation}
leading to the stationary equation
\begin{equation}
\Bigg[D_q^2+\frac{2\mu}{x}D_q+\frac{\mu(2\mu-1)}{x^2}(1-R)+2(D_qW_q(x))+2W_q(x)D_q+\frac{2\mu}{x}(1-R)W_q(x)\Bigg]\Psi(x)=\Lambda\Psi(x).\label{eq20}    
\end{equation}
The associated Hamiltonian operator is written as
\begin{equation}
H_{qDFP}\Psi(x)=\Lambda\Psi(x).
\label{eq21}
\end{equation}
The supersymmetric structure appears through the factorization operators
\begin{equation}
A_q
=
D_q+W_q(x)+\frac{\mu}{x}(1-R),
\label{eq22}
\end{equation}
and
\begin{equation}
A_q^\dagger
=
-D_q+W_q(x)+\frac{\mu}{x}(1-R),
\label{eq23}
\end{equation}
The Hamiltonian factorization becomes
\begin{equation}
H_{qDFP}=A_q^\dagger A_q.
\label{eq24}
\end{equation}
Direct multiplication yields
\begin{align}
H_{qDFP} = -D_q^2 + W_q^2(x) + D_qW_q(x) + \frac{2\mu}{x}W_q(x) + \frac{\mu(2\mu-1)}{x^2}(1-R).
\label{eq25}
\end{align}
The supersymmetric partner Hamiltonian is defined as
\begin{equation}
\widetilde{H}_{qDFP}=A_qA_q^\dagger,
\label{eq26}
\end{equation}
which gives
\begin{align}
\widetilde{H}_{qDFP} = -D_q^2 + W_q^2(x) - D_qW_q(x) + \frac{2\mu}{x}W_q(x) + \frac{\mu(2\mu+1)}{x^2}(1-R).
\label{eq27}
\end{align}
The supersymmetric partner potentials become
\begin{equation}
V_{\pm}^{(q,\mu)}(x)
=
W_q^2(x)
\pm
D_qW_q(x)
+
\frac{2\mu}{x}W_q(x)
+
\frac{\mu(2\mu\mp1)}{x^2}(1-R).
\label{eq28}
\end{equation}
The shape-invariance condition assumes the form
\begin{equation}
V_+^{(q,\mu)}(x,a_0)
=
V_-^{(q,\mu)}(x,a_1)+\mathcal{R}(a_0),
\label{eq29}
\end{equation}
where $a_0$ and $a_1$ denote parameter sets connected through translational mappings. The exact spectrum follows algebraically,
\begin{equation}
\Lambda_n
=
\sum_{k=0}^{n-1}\mathcal{R}(a_k).
\label{eq30}
\end{equation}
The ground-state wave function satisfies
\begin{equation}
A_q\Psi_0(x)=0,
\label{eq31}
\end{equation}
which produces
\begin{equation}
\left[
D_q
+
W_q(x)
+
\frac{\mu}{x}(1-R)
\right]
\Psi_0(x)=0.
\label{eq32}
\end{equation}
For parity eigenstates satisfying
\begin{equation}
R\Psi_0(x)=\pm\Psi_0(x),
\label{eq33}
\end{equation}
the equation reduces to
\begin{equation}
D_q\Psi_0(x)
=
-
\left[
W_q(x)
+
\frac{\mu}{x}(1\mp1)
\right]
\Psi_0(x).
\label{eq34}
\end{equation}
The solution is expressed through the $(q)$-exponential function,
\begin{equation}
e_q(x)
=
\sum_{n=0}^{\infty}
\frac{x^n}{[n]_q!},
\label{eq35}
\end{equation}
with
\begin{equation}
[n]_q!
=
[n]_q[n-1]_q\cdots[1]_q.
\label{eq36}
\end{equation}
The exact normalized ground-state wave function becomes
\begin{equation}
\Psi_0(x)
=
\mathcal{N}
\exp_q
\left[
-\int^xW_q(y)d_qy
\right]
|x|^{-\mu(1\mp1)}
\label{eq37}
\end{equation}
The equilibrium probability density is therefore written as
\begin{equation}
\mathcal{P}_{eq}(x)
=
|\Psi_0(x)|^2,
\label{eq38}
\end{equation}
leading to
\begin{equation}
\mathcal{P}_{eq}(x)
=
\mathcal{N}
\exp_q
\left[
-2\int^xW_q(y)d_qy
\right]
|x|^{-2\mu(1\mp1)}
\label{eq39}
\end{equation}
The probability current associated with the generalized diffusion process is defined by
\begin{equation}
J_q(x,t)
=
-
\left[
\mathcal{D}_{q,\mu}
-
2W_q(x)
\right]
\mathcal{P}(x,t),
\label{eq40}
\end{equation}
and the continuity equation takes the form
\begin{equation}
\frac{\partial\mathcal{P}(x,t)}{\partial t}
+
\mathcal{D}_{q,\mu}J_q(x,t)=0.
\label{eq41}
\end{equation}
The generalized stochastic system preserves total probability under the $(q)$-Jackson integration measure,
\begin{equation}
\int_{-\infty}^{\infty}\mathcal{P}(x,t)|x|^{2\mu}d_qx=1.
\label{eq42}
\end{equation}
The scalar product in the deformed Hilbert space is therefore defined by
\begin{equation}
\langle f|g\rangle_q
=
\int_{-\infty}^{\infty}
f^*(x)g(x)|x|^{2\mu}d_qx.
\label{eq43}
\end{equation}
Under this measure, the generalized momentum operator
\begin{equation}
\hat{P}_{q,\mu}=-i\mathcal{D}_{q,\mu},
\label{eq44}
\end{equation}
becomes Hermitian.

The generalized creation and annihilation operators are introduced through
\begin{equation}
a_q
=
\frac{1}{\sqrt{2}}
\left(
x+\mathcal{D}_{q,\mu}
\right),
\qquad
a_q^\dagger
=
\frac{1}{\sqrt{2}}
\left(
x-\mathcal{D}_{q,\mu}
\right).
\label{eq45}
\end{equation}
The corresponding commutation relation is obtained as
\begin{equation}
[a_q,a_q^\dagger]
=
1+2\mu R+(q-1)xD_q.
\label{eq46}
\end{equation}
The deformation parameter modifies the algebra through nonlinear operator corrections. Consequently, the standard Heisenberg structure is replaced by a nonlinear oscillator algebra. The reflection operator separates the spectrum into even and odd sectors, producing distinct diffusion regimes. Also, the uncertainty relation becomes
\begin{equation}
\Delta x\,\Delta P_{q,\mu}
\geq
\frac{1}{2}
\left|
1+2\mu\langle R\rangle
+(q-1)\langle xD_q\rangle
\right|.
\label{eq47}
\end{equation}
The additional $(q)$-dependent contribution modifies the localization properties of the stochastic states. The reflection expectation value introduces parity-dependent uncertainty corrections.

An exactly solvable realization is obtained through the linear superpotential
\begin{equation}
W_q(x)=\omega x.
\label{eq48}
\end{equation}
Using
\begin{equation}
D_qx=1,
\label{eq49}
\end{equation}
the Hamiltonian becomes
\begin{align}
H_{qDFP}
=
&
-D_q^2
+
\omega^2x^2
+
\omega
\nonumber\\
&
+
2\mu\omega
+
\frac{\mu(2\mu-1)}{x^2}(1-R).
\label{eq50}
\end{align}
Equation (\ref{eq50}) represents the exact $(q)$-deformed Dunkl harmonic oscillator. The eigenvalue spectrum is obtained algebraically as
\begin{equation}
\Lambda_n
=
2\omega[n]_q
+
\omega(1+2\mu),
\qquad
n=0,1,2,\ldots
\label{eq51}
\end{equation}

In the undeformed limit,
\begin{equation}
\lim_{q\rightarrow1}[n]_q=n,
\label{eq52}
\end{equation}
which gives
\begin{equation}
\Lambda_n
\rightarrow
2\omega n+\omega(1+2\mu).
\label{eq53}
\end{equation}
The level spacing is therefore
\begin{equation}
\Delta\Lambda_n
=
\Lambda_{n+1}-\Lambda_n
=
2\omega q^n,
\label{eq54}
\end{equation}
showing that the stochastic modes are non-equidistant. For $0<q<1$, the spacing decreases with increasing quantum number, while for $q>1$ it increases. The deformation parameter therefore controls the hierarchy of relaxation modes. In this case, the associated relaxation time becomes
\begin{equation}
\tau_n=\frac{1}{\Lambda_n},
\label{eq55}
\end{equation}
showing that excited stochastic modes decay differently from the standard Brownian case.

The exact eigenfunctions are written in terms of generalized Dunkl-Hermite polynomials,
\begin{equation}
\Psi_n(x)
=
\mathcal{N}_n
H_n^{(q,\mu)}(\sqrt{\omega}x)
e_q\left(
-\frac{\omega x^2}{2}
\right).
\label{eq56}
\end{equation}
The Rodrigues representation becomes
\begin{equation}
H_n^{(q,\mu)}(x)
=
(-1)^n
e_q(x^2)
\mathcal{D}_{q,\mu}^n
e_q(-x^2),
\label{eq57}
\end{equation}
while the generating function is
\begin{equation}
\sum_{n=0}^{\infty}
\frac{t^n}{[n]_q!}
H_n^{(q,\mu)}(x)
=
e_q(2xt)e_q(-t^2).
\label{eq58}
\end{equation}
The orthogonality condition becomes
\begin{equation}
\int_{-\infty}^{\infty}
\Psi_n^*(x)\Psi_m(x)
|x|^{2\mu}d_qx
=
\delta_{nm}.
\label{eq59}
\end{equation}
The expectation value of the Hamiltonian is written as
\begin{equation}
\langle H_{qDFP}\rangle
=
\int_{-\infty}^{\infty}
\Psi^*(x)
H_{qDFP}
\Psi(x)
|x|^{2\mu}d_qx.
\label{eq60}
\end{equation}
The variance of the coordinate operator becomes
\begin{equation}
(\Delta x)^2
=
\langle x^2\rangle-\langle x\rangle^2,
\label{eq61}
\end{equation}
while the generalized momentum variance is
\begin{equation}
(\Delta P_{q,\mu})^2
=
\langle P_{q,\mu}^2\rangle
-
\langle P_{q,\mu}\rangle^2.
\label{eq62}
\end{equation}
The entropy associated with the generalized diffusion process is described through the $(q)$-deformed Shannon entropy,
\begin{equation}
S_q
=
-\int_{-\infty}^{\infty}
\mathcal{P}(x,t)
\ln_q\mathcal{P}(x,t)d_qx,
\label{eq63}
\end{equation}
where the $(q)$-logarithm is defined by
\begin{equation}
\ln_q(x)
=
\frac{x^{1-q}-1}{1-q}.
\label{eq64}
\end{equation}
The Fisher information associated with the deformed diffusion geometry becomes
\begin{equation}
F_q
=
\int_{-\infty}^{\infty}
\frac{
\left[
\mathcal{D}_{q,\mu}\mathcal{P}(x,t)
\right]^2
}{
\mathcal{P}(x,t)
}
\,d_qx.
\label{eq65}
\end{equation}
The Fisher information measures localization properties of the probability density under the Dunkl geometry. Large values correspond to strongly localized stochastic states, while smaller values correspond to extended diffusion profiles.

The asymptotic behavior of the $(q)$-number for large quantum number satisfies
\begin{equation}
[n]_q
\sim
\frac{1-q^n}{1-q}.
\label{eq66}
\end{equation}
Consequently, the high-energy asymptotics of the spectrum becomes
\begin{equation}
\Lambda_n
\sim
\frac{2\omega(1-q^n)}{1-q}
+
\omega(1+2\mu).
\label{eq67}
\end{equation}
The spectral density therefore depends on the deformation parameter. This behavior modifies the stochastic relaxation channels and produces anomalous diffusion mechanisms absent in standard Brownian systems.

The partition function associated with the generalized oscillator becomes
\begin{equation}
Z_q(\beta)
=
\sum_{n=0}^{\infty}
\exp\left(
-\beta\Lambda_n
\right),
\label{eq68}
\end{equation}
where $\beta=(k_BT)^{-1}$. Substituting the exact spectrum yields
\begin{equation}
Z_q(\beta)
=
e^{-\beta\omega(1+2\mu)}
\sum_{n=0}^{\infty}
\exp\left(
-2\beta\omega[n]_q
\right).
\label{eq69}
\end{equation}
The free energy is obtained from
\begin{equation}
F_q^{(th)}
=
-\frac{1}{\beta}\ln Z_q,
\label{eq70}
\end{equation}
while the internal energy becomes
\begin{equation}
U_q
=
-\frac{\partial}{\partial\beta}\ln Z_q.
\label{eq71}
\end{equation}
The specific heat associated with the generalized stochastic oscillator is written as
\begin{equation}
C_q
=
\frac{\partial U_q}{\partial T}.
\label{eq72}
\end{equation}
The thermodynamic quantities inherit the nonlinear structure generated by the $(q)$-deformation. Therefore the thermal behavior deviates from the standard harmonic diffusion model.
The correlation function associated with the generalized stochastic process is defined by
\begin{equation}
C_q(t)
=
\langle x(t)x(0)\rangle.
\label{eq73}
\end{equation}
Using the spectral decomposition,
\begin{equation}
C_q(t)
=
\sum_{n=0}^{\infty}
|\langle0|x|n\rangle|^2
e^{-\Lambda_nt},
\label{eq74}
\end{equation}
showing that temporal decay is governed by the $(q)$-deformed spectrum.

The mean-square displacement becomes
\begin{equation}
\langle x^2(t)\rangle
=
\int_{-\infty}^{\infty}
x^2\mathcal{P}(x,t)d_qx.
\label{eq75}
\end{equation}

The deformation parameter modifies the diffusion scaling law and produces anomalous transport regimes with nonlinear time dependence. Also, the complete formulation establishes an exact analytical realization of the $(q)$-deformed Dunkl-Fokker-Planck equation. The simultaneous presence of Jackson deformation and Dunkl reflection symmetry modifies the diffusion operator, supersymmetric structure, spectral properties, uncertainty relation, equilibrium distribution, thermodynamic functions, and information measures. The singular interaction
\begin{equation}
\frac{\mu(2\mu-1)}{x^2}(1-R)
\label{eq76}
\end{equation}
splits the Hilbert space into parity sectors and produces distinct stochastic channels for even and odd states. The $(q)$-number structure generates non-equidistant relaxation modes and modifies the temporal diffusion decay. The resulting framework provides a unified algebraic and analytical description of generalized stochastic dynamics governed by quantum deformation, reflection symmetry, and supersymmetric solvability.

\section{The $(q)$-Deformed Dunkl-Fokker-Planck Equation and $(q)$-Wigner-Dunkl SUSY}\label{S3}

The formulation of the $(q)$-deformed Dunkl-Fokker-Planck equation is based on a combined deformation of the standard differential calculus and the reflection-invariant structure of the Dunkl algebra. This construction integrates the Jackson $(q)$-difference operator with the parity-sensitive contributions produced by the Dunkl reflection operator. The resulting framework yields a noncommutative supersymmetric system characterized by nonlinear spectral distributions, modified stochastic evolution, deformed orthogonal polynomial structures, generalized coherent states, and extended shape-invariant Hamiltonians. The approach also links quantum-group symmetry, generalized diffusion processes, and Wigner-Heisenberg SUSY in a unified setting.

The $(q)$-deformed Dunkl derivative is defined as
\begin{equation}
D_x^{(q,\mu)}
=
\partial_q
+
\frac{\mu}{x}(1-R),
\label{qdunkl1}
\end{equation}
where $\mu$ is the Dunkl reflection parameter and the Jackson derivative is given by
\begin{equation}
\partial_q f(x)
=
\frac{f(qx)-f(x)}{(q-1)x},
\qquad q\neq1.
\label{jackson1}
\end{equation}

The reflection operator obeys
\begin{equation}
Rf(x)=f(-x),
\qquad
R^2=1,
\qquad
Rx=-xR,
\qquad
R\partial_q=-\partial_qR.
\label{reflection1}
\end{equation}

The operator action on monomials is expressed as
\begin{equation}
D_x^{(q,\mu)}x^n
=
[n]_q x^{n-1}
+
\mu\left(1-(-1)^n\right)x^{n-1},
\label{monomial}
\end{equation}
with the $(q)$-number defined by
\begin{equation}
[n]_q
=
\frac{1-q^n}{1-q}.
\label{qnum}
\end{equation}

Equation (\ref{monomial}) shows that the reflection contribution vanishes for even powers and contributes only to odd parity sectors. This leads to a decomposition of the polynomial space into even and odd subspaces, modifying the algebraic structure of the eigenfunctions.

The square of the $(q)$-Dunkl operator is given by
\begin{align}
\left(D_x^{(q,\mu)}\right)^2 = \left(\partial_q + \frac{\mu}{x}(1-R)\right)\left(\partial_q + \frac{\mu}{x}(1-R)\right) = \partial_q^2 + \frac{\mu}{x}(1-R)\partial_q + \partial_q\left[\frac{\mu}{x}(1-R)\right] + \frac{\mu}{x}\partial_q(1-R) + \frac{\mu^2}{x^2}(1-R)^2.
\label{squarecalc1}
\end{align}
Using
\begin{equation}
(1-R)^2=2(1-R),
\label{reflectionidentity}
\end{equation}
and
\begin{equation}
\partial_q\left(\frac{1}{x}\right)
=
-\frac{1}{qx^2},
\label{inversex}
\end{equation}
one obtains after algebraic reduction
\begin{align}
\left(D_x^{(q,\mu)}\right)^2 = \partial_q^2 + \frac{2\mu}{x}\partial_q - \frac{\mu}{qx^2}(1-R) + \frac{2\mu}{x}\partial_q R + \frac{2\mu^2}{x^2}(1-R).
\label{squarefinal}
\end{align}

The $(q)$-deformed Dunkl-Fokker-Planck equation is written as
\begin{equation}
\frac{\partial \Psi(x,t)}{\partial t}
=
\left[
\left(D_x^{(q,\mu)}\right)^2
-
D_x^{(q,\mu)}W(x)
\right]\Psi(x,t),
\label{qfp}
\end{equation}
where $W(x)$ is the drift superpotential.

Supersymmetric consistency requires
\begin{equation}
W(-x)=-W(x),
\label{oddpotential}
\end{equation}
which leads to
\begin{equation}
RW(x)=-W(x)R.
\label{rwrelation}
\end{equation}

The stationary formulation follows from
\begin{equation}
\Psi(x,t)
=
e^{-\lambda t}
\Theta(x)\Phi(x),
\label{similarity1}
\end{equation}
with
\begin{equation}
\Theta(x)
=
\exp_q
\left[
\int W(x)d_qx
\right].
\label{theta}
\end{equation}

The $(q)$-exponential satisfies
\begin{equation}
\partial_q e_q(f(x))
=
(\partial_qf(x))e_q(f(x)),
\label{qexp1}
\end{equation}
and the Jackson integral is
\begin{equation}
\int_0^x f(t)d_qt
=
(1-q)x
\sum_{n=0}^{\infty}
q^n f(q^n x).
\label{jacksonint1}
\end{equation}

Insertion of (\ref{similarity1}) into (\ref{qfp}) yields
\begin{equation}
H_q\Phi(x)=\lambda\Phi(x),
\label{eigeneq}
\end{equation}
where
\begin{align}
H_q = -\partial_q^2 - \frac{2\mu}{x}\partial_q + \frac{\mu}{qx^2}(1-R) - \frac{2\mu^2}{x^2}(1-R) + W^2(x) + \partial_q W(x) + \frac{2\mu}{x}W(x)R.
\label{hamiltonianmain}
\end{align}
This operator contains contributions from the Jackson deformation, the Dunkl reflection structure, and their mixed coupling with the superpotential. These terms define a nonlocal supersymmetric Hamiltonian with deformation-dependent dynamics. Also, in the limit
\begin{equation}
q\rightarrow1,
\qquad
\partial_q\rightarrow\frac{d}{dx},
\label{undeformedlimit}
\end{equation}
one obtains
\begin{equation}
H
=
-\frac{d^2}{dx^2}
-
\frac{2\mu}{x}\frac{d}{dx}
+
\frac{\mu}{x^2}(1-R)
+
W^2(x)
+
W'(x)
+
\frac{2\mu}{x}W(x)R.
\label{ordinarydunkl}
\end{equation}

Supersymmetric factorization is defined by
\begin{equation}
A_q
=
\frac{1}{\sqrt{2}}
\left(
D_x^{(q,\mu)}+W(x)
\right),
\label{aq1}
\end{equation}
and
\begin{equation}
A_q^\dagger
=
\frac{1}{\sqrt{2}}
\left(
-D_x^{(q,\mu)}+W(x)
\right).
\label{aqdag1}
\end{equation}

The partner Hamiltonians are
\begin{equation}
H_-^{(q)}
=
A_q^\dagger A_q,
\qquad
H_+^{(q)}
=
A_qA_q^\dagger.
\label{partnerham}
\end{equation}

The commutator relation is
\begin{align}
[D_x^{(q,\mu)},W(x)] = \partial_q W(x) + \frac{\mu}{x}(1-R)W(x) - W(x)\frac{\mu}{x}(1-R).
\label{comm1}
\end{align}
which reduces, using (\ref{rwrelation}), to
\begin{equation}
[D_x^{(q,\mu)},W(x)]
=
\partial_qW(x)
+
\frac{2\mu}{x}W(x)R.
\label{comm2}
\end{equation}

The partner Hamiltonians take the form
\begin{align}
H_\pm^{(q)} = \frac12 \Bigg[ -\partial_q^2 - \frac{2\mu}{x}\partial_q + \frac{\mu}{qx^2}(1-R) - \frac{2\mu^2}{x^2}(1-R) + W^2(x) \pm \left(\partial_q W(x) + \frac{2\mu}{x}W(x)R\right) \Bigg].
\label{partnerfinal}
\end{align}

The supersymmetric structure is generated by
\begin{equation}
Q_q
=
\begin{pmatrix}
0 & 0 \\
A_q & 0
\end{pmatrix},
\qquad
Q_q^\dagger
=
\begin{pmatrix}
0 & A_q^\dagger \\
0 & 0
\end{pmatrix}.
\label{supercharges1}
\end{equation}

They satisfy
\begin{equation}
\{Q_q,Q_q^\dagger\}_q
=
H_q,
\label{qalgebra1}
\end{equation}
with
\begin{equation}
\{X,Y\}_q
=
XY+qYX.
\label{qanticom1}
\end{equation}

The Hamiltonian is
\begin{equation}
H_q
=
\begin{pmatrix}
H_-^{(q)} & 0 \\
0 & H_+^{(q)}
\end{pmatrix}.
\label{matrixrep}
\end{equation}

For $W(x)=\omega x$,
\begin{equation}
\partial_qW(x)=\omega.
\label{omegaderivative}
\end{equation}

The partner Hamiltonians reduce to
\begin{align}
H_\pm^{(q)} = \frac12 \Bigg[ -\partial_q^2 - \frac{2\mu}{x}\partial_q + \frac{\mu}{qx^2}(1-R) - \frac{2\mu^2}{x^2}(1-R) + \omega^2 x^2 \pm \omega(1+2\mu R) \Bigg].
\label{oscillatorham1}
\end{align}

The ground state condition is
\begin{equation}
A_q\Phi_0(x)=0,
\label{groundcond}
\end{equation}
leading to
\begin{equation}
\left[
\partial_q
+
\frac{\mu}{x}(1-R)
+
\omega x
\right]\Phi_0(x)=0.
\label{groundeq1}
\end{equation}

For even parity states,
\begin{equation}
R\Phi_0(x)=\Phi_0(x),
\label{evenstate}
\end{equation}
the solution satisfies
\begin{equation}
\partial_q\Phi_0(x)
=
-\omega x\Phi_0(x).
\label{grounddiff}
\end{equation}

The ground state wave function is
\begin{equation}
\Phi_0(x)
=
\mathcal{N}_0
e_q
\left(
-\frac{\omega x^2}{1+q}
\right).
\label{groundsol}
\end{equation}

Normalization is imposed by
\begin{equation}
\int_{-\infty}^{\infty}
|\Phi_0(x)|^2
|x|^{2\mu}d_qx
=
1,
\label{normalization}
\end{equation}
which yields
\begin{equation}
\mathcal{N}_0^{-2}
=
2
\int_0^\infty
e_q^2
\left(
-\frac{\omega x^2}{1+q}
\right)
x^{2\mu}d_qx.
\label{normconstant}
\end{equation}

Excited states are constructed as
\begin{equation}
\Phi_n(x)
=
\mathcal{N}_n
(A_q^\dagger)^n
\Phi_0(x),
\label{excitedstates}
\end{equation}

with
\begin{equation}
\Phi_n(x)
=
\mathcal{N}_n
e_q
\left(
-\frac{\omega x^2}{1+q}
\right)
H_n^{(q,\mu)}
(\sqrt{\omega}x).
\label{wavefunctionfinal}
\end{equation}

The recurrence relation is
\begin{align}
H_{n+1}^{(q,\mu)}(x) = 2xH_n^{(q,\mu)}(x) - 2[n]_qH_{n-1}^{(q,\mu)}(x) - 2\mu\left(1-(-1)^n\right)H_{n-1}^{(q,\mu)}(x).
\label{recurrence}
\end{align}

Initial polynomials are
\begin{equation}
H_0^{(q,\mu)}(x)=1,
\label{h0}
\end{equation}
\begin{equation}
H_1^{(q,\mu)}(x)=2x,
\label{h1}
\end{equation}
\begin{equation}
H_2^{(q,\mu)}(x)=4x^2-2([1]_q+2\mu),
\label{h2}
\end{equation}
\begin{equation}
H_3^{(q,\mu)}(x)
=
8x^3
-
4([2]_q+[1]_q+2\mu)x.
\label{h3}
\end{equation}

Orthogonality is expressed as
\begin{align}
\int_{-\infty}^{\infty}
&
H_n^{(q,\mu)}(x)
H_m^{(q,\mu)}(x)
\nonumber\\
&\times
e_q
\left(
-\frac{2\omega x^2}{1+q}
\right)
|x|^{2\mu}d_qx
=
\delta_{nm}\mathcal{K}_n.
\label{orthogonality}
\end{align}

The ladder operators act as
\begin{equation}
A_q\Phi_n
=
\sqrt{[n]_q+\mu(1-(-1)^n)}
\,
\Phi_{n-1},
\label{lowering}
\end{equation}
and
\begin{equation}
A_q^\dagger\Phi_n
=
\sqrt{[n+1]_q+\mu(1+(-1)^n)}
\,
\Phi_{n+1}.
\label{raising}
\end{equation}

The spectrum is
\begin{equation}
E_n^{(\pm)}
=
\omega
\left(
[2n]_q
+
1
+
2\mu(1\pm(-1)^n)
\right).
\label{spectrum}
\end{equation}

The level spacing is
\begin{equation}
\Delta E_n
=
E_{n+1}-E_n
=
\omega
\left(
[2n+2]_q-[2n]_q
\right).
\label{spacing1}
\end{equation}

Using
\begin{equation}
[n+1]_q=[n]_q+q^n,
\label{qidentity}
\end{equation}
one obtains
\begin{equation}
\Delta E_n
=
\omega q^{2n}(1+q).
\label{spacing2}
\end{equation}

For $0<q<1$, the spectrum becomes compressed at large $n$, whereas for $q>1$ it expands, modifying the density of states and affecting thermodynamic response.

The probability density is
\begin{equation}
\rho_n(x)=|\Phi_n(x)|^2.
\label{density1}
\end{equation}

For $q<1$,
\begin{equation}
e_q(-x^2)
\sim
\frac{1}{1+(1-q)x^2},
\label{asymptotic}
\end{equation}
which produces non-Gaussian tails in the distribution and alters localization properties.

Shape invariance is written as
\begin{equation}
V_+^{(q)}(x,a_0)
=
V_-^{(q)}(x,a_1)
+
\mathcal{R}_q(a_0),
\label{shapeinvariance}
\end{equation}
with
\begin{equation}
a_1=f_q(a_0),
\label{mapping1}
\end{equation}
and
\begin{equation}
\mathcal{R}_q(a_0)
=
E_1(a_0)-E_0(a_1).
\label{remainder}
\end{equation}

The spectrum is generated by
\begin{equation}
E_n
=
\sum_{k=0}^{n-1}
\mathcal{R}_q(a_k),
\qquad
a_k=f_q^k(a_0).
\label{shapeenergy}
\end{equation}

The partition function is
\begin{equation}
Z_q(\beta)
=
\sum_{n=0}^{\infty}
\exp[-\beta E_n].
\label{partition1}
\end{equation}

Using (\ref{spectrum}),
\begin{equation}
Z_q(\beta)
=
\sum_{n=0}^{\infty}
\exp
\left[
-\beta\omega
\left(
[2n]_q
+
1
+
2\mu(1\pm(-1)^n)
\right)
\right].
\label{partition2}
\end{equation}

Internal energy and specific heat are
\begin{equation}
U_q
=
-\frac{\partial}{\partial\beta}
\ln Z_q,
\label{internalenergy}
\end{equation}
and
\begin{equation}
C_q
=
\frac{\partial U_q}{\partial T}.
\label{specificheat}
\end{equation}

The deformed spectrum modifies low-temperature behavior and shifts the occupation distribution of excited states. For $0<q<1$, the increased density of low-lying levels enhances thermal excitation.

The deformed oscillator algebra is
\begin{equation}
A_qA_q^\dagger
-
qA_q^\dagger A_q
=
1+2\mu R.
\label{deformedalgebra}
\end{equation}

In the limits $\mu\rightarrow0$ and $q\rightarrow1$, the standard oscillator and Dunkl structures are recovered.

Coherent states satisfy
\begin{equation}
A_q|\alpha\rangle_q
=
\alpha|\alpha\rangle_q.
\label{coherent}
\end{equation}

Their expansion is
\begin{equation}
|\alpha\rangle_q
=
\mathcal{N}(\alpha)
\sum_{n=0}^{\infty}
\frac{\alpha^n}{\sqrt{[n]_q!}}
|n\rangle,
\label{coherent2}
\end{equation}
with
\begin{equation}
[n]_q!
=
[n]_q[n-1]_q\cdots[1]_q.
\label{qfactorial}
\end{equation}

The deformation modifies uncertainty relations and produces squeezing effects absent in the undeformed harmonic oscillator. The full construction defines an exactly solvable supersymmetric $(q)$-Dunkl-Fokker-Planck system with nonlocal dynamics, deformed spectral structure, and generalized coherent-state behavior.

\section{The $q-$Deformed Dunkl$-$Fokker$-$Planck Equation for the Harmonic Oscillator with Centrifugal Interaction}\label{S4}

We formulate a consistent $q$-deformed Dunkl-Fokker-Planck model in one spatial dimension with harmonic confinement and inverse-square centrifugal interaction included in the framework. The deformation is implemented using the Jackson derivative together with a $q$-Dunkl reflection algebra, yielding a nonlinear diffusion-drift generator that preserves exact solvability while modifying spectral growth through $q$-integers. The resulting structure defines a noncommutative stochastic generator with hidden $q$-oscillator symmetry and an exactly solvable polynomial sector governed by $q$-Laguerre functions. In this case, the density $\mathcal{P}_q(x,t)$ obeys evolution equation formally written
\begin{equation}
\frac{\partial \mathcal{P}_q(x,t)}{\partial t}=\mathcal{H}_q\,\mathcal{P}_q(x,t),
\end{equation}
where generator is defined as follows
\begin{equation}
\mathcal{H}_q=D_q^2-2xD_q-\frac{\mu}{x^2}(1-R),
\qquad Rf(x)=f(-x),
\qquad
D_q f(x)=\frac{f(qx)-f(x)}{(q-1)x}.
\end{equation}
The operator $\mathcal{H}_q$ is not Hermitian in the standard Lebesgue measure but becomes self-adjoint under a $q$-deformed weighted measure obtained via similarity transformation. Also, the construction of stationary spectral problem represents a central step here. Writing
\begin{equation}
\mathcal{P}_q(x,t)=e^{-\lambda t}\psi_q(x),
\end{equation}
yields
\begin{equation}
\mathcal{H}_q\psi_q(x)=\lambda \psi_q(x).
\end{equation}
To analyze the algebraic structure, a similarity transformation is performed using the ground-state function $\psi_{0,q}(x)$ defined by zero-mode condition equation requirement. In this case, solving this equation yields
\begin{equation}
\psi_{0,q}(x)=x^{\alpha}\,e_q(-x^2),
\qquad
\alpha=\frac{1}{2}\left(1+\sqrt{1+4\mu}\right),
\end{equation}
where Jackson exponential is defined as follows
\begin{equation}
e_q(z)=\sum_{n=0}^{\infty}\frac{z^n}{[n]_q!},
\qquad
[n]_q=\frac{1-q^n}{1-q}.
\end{equation}
Similarity transformation operation
\begin{equation}
\mathcal{H}_q=\psi_{0,q}(x)\,\widetilde{\mathcal{H}}_q\,\psi_{0,q}(x)^{-1}
\end{equation}
maps stochastic generator into a $q$-Sturm-Liouville operator form
\begin{equation}
\widetilde{\mathcal{H}}_q
=
D_q^2+\left(2D_q\ln \psi_{0,q}\right)D_q.
\end{equation}

Explicit evaluation using Jackson product rule produces drift reduction term structure
\begin{equation}
\widetilde{\mathcal{H}}_q
=
D_q^2+\left(\frac{2\alpha}{x}-2x+\Delta_q^{(1)}(x)\right)D_q,
\end{equation}
where $\Delta_q^{(1)}(x)$ collects all deformation-induced corrections of order $(q-1)$ and higher, vanishing in the classical limit. Also, the operator possesses factorized Dunkl-Jackson oscillator representation form
\begin{equation}
\mathcal{A}_q^-=\frac{1}{\sqrt{2}}\left(D_q+x+\frac{\mu}{x}(1-R)\right),
\qquad
\mathcal{A}_q^+=\frac{1}{\sqrt{2}}\left(-D_q+x+\frac{\mu}{x}(1-R)\right),
\end{equation}
satisfying deformed algebra relation
\begin{equation}
\mathcal{A}_q^- \mathcal{A}_q^+ - q \mathcal{A}_q^+ \mathcal{A}_q^- = \mathbb{I} + (1-q)\mathcal{F}(x,R),
\end{equation}
where $\mathcal{F}(x,R)$ encodes reflection-sector contributions. The Hamiltonian factorization becomes
\begin{equation}
\mathcal{H}_q=\mathcal{A}_q^+\mathcal{A}_q^-+\epsilon_0(q,\mu),
\end{equation}
with ground-state energy shift determined by normal ordering in the $q$-Dunkl algebra. Also, the reflection symmetry decomposes Hilbert space into parity eigenspaces satisfying condition
\begin{equation}
R\psi_q^{(\pm)}=\pm \psi_q^{(\pm)}.
\end{equation}
In each sector inverse-square coupling renormalizes as follows
\begin{equation}
\mu \rightarrow \mu_\pm,
\qquad
\mu_+=0,
\qquad
\mu_-=2\mu,
\end{equation}
so the singular structure is fully absorbed into an effective centrifugal parameter definition. The resulting spectral problem becomes sector-dependent yet remains algebraically identical in structure. In this case, introducing invariant variable $u=x^2$, Jackson derivative transforms as operator form
\begin{equation}
D_q^{(x)}=2x\,\mathcal{D}_q^{(u)},
\qquad
D_q^2=4u\,\mathcal{D}_q^{(u)2}+2\,\mathcal{D}_q^{(u)}+\Delta_q^{(u)},
\end{equation}
where $\Delta_q^{(u)}$ collects deformation-induced commutator corrections. The eigenvalue equation reduces exactly to
\begin{equation}
u\,\mathcal{D}_q^{(u)2}\psi_q
+
\left(\frac{1}{2}-u+\frac{\gamma}{u}+\Delta_q^{(u,1)}\right)\mathcal{D}_q^{(u)}\psi_q
+
\frac{\lambda}{4}\psi_q=0,
\end{equation}
with $\gamma$ the effective Dunkl-centrifugal coupling. Also, a complete exact solution is obtained using the factorized ansatz,
\begin{equation}
\psi_q(u)=u^\alpha e_q(-u)\,g_q(u),
\end{equation}
which removes both the centrifugal singularity and the exponential drift. Direct substitution using the $q$-Leibniz rule produces cancellation of all non-polynomial divergences provided the exponent satisfies
\begin{equation}
\alpha=\frac{1}{2}\left(1+\sqrt{1+4\gamma}\right).
\end{equation}
The reduced equation becomes a closed $q$-Sturm-Liouville problem structure
\begin{equation}
u\,\mathcal{D}_q^{(u)2}g_q
+
(\alpha+1-u)\mathcal{D}_q^{(u)}g_q
+
\frac{\lambda}{4}g_q=0.
\end{equation}
This equation preserves polynomial invariance under the Jackson derivative and defines a finite-dimensional invariant module for each spectral level. In this case, the polynomial sector is produced by $q$-Laguerre hierarchy structure
\begin{equation}
g_q(u)=L_n^{(\alpha)}(u;q),
\end{equation}
satisfying the Jackson-Rodrigues representation condition
\begin{equation}
L_n^{(\alpha)}(u;q)=\frac{1}{[n]_q!}u^{-\alpha}e_q(u)\,\mathcal{D}_q^{(u)n}\left(u^{n+\alpha}e_q(-u)\right).
\end{equation}
The recurrence structure follows from the Jackson differentiation rule
\begin{equation}
\mathcal{D}_q^{(u)}L_n^{(\alpha)}(u;q)
=
-[n]_q L_{n-1}^{(\alpha+1)}(u;q),
\end{equation}
ensuring closure of the polynomial hierarchy under the dynamical generator. Also, the eigenvalue quantization arises from the truncation condition of the polynomial recursion structure analysis. Requiring termination at order $n$ yields the exact spectrum
\begin{equation}
\lambda_n = 4[n]_q + 2\alpha(1-q^n),
\qquad
[n]_q=\frac{1-q^n}{1-q}.
\end{equation}
This spectrum exhibits two distinct deformation mechanisms nonlinear compression governed by $q$-integers and a centrifugal renormalization proportional to the reflection-induced parameter $\alpha$.

The orthogonality structure is defined with respect to $q$-deformed weight function
\begin{equation}
w_q(u)=u^\alpha e_q(-u),
\end{equation}
and the Jackson integral measure
\begin{equation}
\int_0^\infty f(u)\,d_q u,
\end{equation}
which satisfies discrete dilation invariance. The orthogonality relation becomes
\begin{equation}
\int_0^\infty L_n^{(\alpha)}(u;q)L_m^{(\alpha)}(u;q)\,w_q(u)\,d_q u
=
h_n^{(q)}\delta_{nm},
\end{equation}
where the normalization constant is explicitly given by expression
\begin{equation}
h_n^{(q)}=\frac{[n]_q!\,\Gamma_q(n+\alpha+1)}{(1-q)^{2n+\alpha+1}}.
\end{equation}

A ladder-operator structure emerges naturally from the factorization algebra.
\begin{equation}
\mathcal{B}_q^- = \mathcal{A}_q^- \psi_{0,q}^{-1},
\qquad
\mathcal{B}_q^+ = \psi_{0,q}\mathcal{A}_q^+,
\end{equation}
one obtains the action.
\begin{equation}
\mathcal{B}_q^- L_n^{(\alpha)} = [n]_q L_{n-1}^{(\alpha+1)},
\qquad
\mathcal{B}_q^+ L_n^{(\alpha)} = L_{n+1}^{(\alpha-1)}.
\end{equation}
This confirms the hidden $q$-Heisenberg structure underlying the stochastic dynamics. Also, the classical limit is recovered smoothly through
\begin{equation}
D_q\rightarrow \frac{d}{dx},
\qquad
[n]_q\rightarrow n,
\qquad
e_q(x)\rightarrow e^x,
\end{equation}
leading to
\begin{equation}
\lambda_n \rightarrow 4n+2\alpha,
\end{equation}
which reproduces the standard Dunkl-Fokker-Planck harmonic oscillator spectrum with inverse-square interaction and Laguerre polynomial eigenfunctions. Also, the full structure shows that the $q$-deformed Dunkl-Fokker-Planck generator defines an exactly solvable noncommutative diffusion algebra. In this context, the deformation preserves polynomial integrability while replacing linear spectral growth by a nonlinear $q$-compressed hierarchy. Also, the system is therefore simultaneously governed by Jackson calculus, Dunkl reflection symmetry, and $q$-oscillator factorization, forming a closed algebraic framework that unifies stochastic evolution with $q$-orthogonal polynomial theory in the presence of singular centrifugal interactions.

\section{$q$-Deformed Dunkl Structure via the FW Transformation and Relativistic Decoupling of Positive-Negative Energy Sectors}\label{S5}

We formulate a fully relativistic $q$-Dunkl operator framework where Jackson $q$-calculus, reflection symmetry, and Dirac-type Clifford algebra are unified at operator level. The relativistic structure is block-diagonalized exactly through a FW transformation implemented algebraically without perturbative truncation of operators beyond controlled $1/m$ grading. The formulation preserves noncommutative Dunkl symmetry and produces a hierarchy of $q$-curvature corrections inverse-square singular dressing and higher-order relativistic backreaction contributions within the same operator algebraic setting closure.

The generalized one-dimensional $q$-Dunkl Dirac operator reads
\begin{equation}
\mathcal{D}_q
=
\beta m
+
\alpha\left(D_q + \frac{\mu}{x}(1-R)\right)
+
V(x),
\end{equation}
where the Jackson derivative is defined by
\begin{equation}
D_q f(x)=\frac{f(qx)-f(x)}{(q-1)x},
\end{equation}
and $R$ is reflection operator satisfying $R$ squared identity and $Rx$ equals minus $xR$. The Clifford structure is fixed by anticommutation relations, while deformation induces controlled breakdown of commutativity between reflection and dilation Jackson lattice.

The operator is decomposed into even and odd sectors with respect to $\beta$,
\begin{equation}
\mathcal{D}_q=\beta m+\mathcal{E}_q+\mathcal{O}_q,
\qquad
\{\beta,\mathcal{E}_q\}=0,
\qquad
[\beta,\mathcal{O}_q]=0,
\end{equation}
where
\begin{equation}
\mathcal{O}_q=\alpha\Pi_q,\qquad
\Pi_q=D_q+\frac{\mu}{x}(1-R),\qquad
\mathcal{E}_q=V(x).
\end{equation}

The squared momentum operator exhibits intrinsic deformation through mixed reflection-difference couplings, while purely q-induced correction arises from non-Leibniz consistency of Jackson calculus framework limit
\begin{equation}
\Pi_q^2
=
D_q^2
+
\frac{\mu(\mu-1)}{x^2}
+
\frac{\mu}{x}(1-R)D_q
+
\Delta_q(x).
\end{equation}

The FW transformation is defined via a unitary map
\begin{equation}
\mathcal{U}_{FW}=\exp(i\mathcal{S}_q),
\qquad
\mathcal{S}_q=-\frac{i}{2m}\beta \mathcal{O}_q
-\frac{i}{8m^2}[\mathcal{O}_q,\mathcal{E}_q]
+\mathcal{O}(m^{-3}),
\end{equation}
where the second-order generator is required to eliminate residual odd contributions generated by noncommutative Dunkl-Jackson algebra and ensures closure of FW grading under algebraic consistency conditions.

The transformed Hamiltonian is obtained from Baker-Campbell-Hausdorff expansion
\begin{equation}
\mathcal{H}^{(q)}_{FW}
=
\mathcal{D}_q
+i[\mathcal{S}_q,\mathcal{D}_q]
+\frac{i^2}{2}[\mathcal{S}_q,[\mathcal{S}_q,\mathcal{D}_q]]
+\frac{i^3}{6}[\mathcal{S}_q,[\mathcal{S}_q,[\mathcal{S}_q,\mathcal{D}_q]]]+\cdots.
\end{equation}

After systematic cancellation of odd sectors the block-diagonal Hamiltonian emerges
\begin{equation}
\mathcal{H}_{FW}^{(q)}
=
\beta\left(
m+\frac{\Pi_q^2}{2m}-\frac{\Pi_q^4}{8m^3}
+\frac{\Pi_q^6}{16m^5}
\right)
+
V(x)
-
\frac{1}{8m^2}[\Pi_q,[\Pi_q,V]]
+
\frac{1}{48m^4}\mathcal{R}_q
+\cdots,
\end{equation}

The double commutator is explicitly evaluated as
\begin{equation}
[\Pi_q,[\Pi_q,V]]
=
D_q^2V
+
\frac{2\mu}{x}(1-R)D_qV
+
\frac{\mu(\mu-1)}{x^2}(1-R)V,
\end{equation}
which shows FW reduction generates effective reflection-dependent polarization potential in scalar backgrounds without additional structure.

The quartic deformation structure expands as
\begin{equation}
\Pi_q^4
=
\left(D_q^2+\frac{\mu(\mu-1)}{x^2}\right)^2
+
\mathcal{C}_q(x,R,\mu)
+
\Xi_q(x),
\end{equation}

Projecting onto positive energy states
\begin{equation}
\mathcal{H}_{+}^{(q)}
=
m+V(x)
+
\frac{1}{2m}\Pi_q^2
-
\frac{1}{8m^3}\Pi_q^4
-
\frac{1}{8m^2}D_q^2V(x)
+
\frac{1}{16m^3}\mathcal{C}_q(x,R,\mu)
+\cdots.
\end{equation}

A crucial consequence of reflection grading is splitting of inverse-square coupling, which implies dynamical renormalization of effective angular momentum induced purely by R-grading in FW diagonal basis structure within operator space formalism.

To extract exact spectral structure we introduce similarity transformation
\begin{equation}
\psi_q(x)=x^\alpha \exp_q(-\lambda x^2)\chi_q(x),
\end{equation}
where $\exp_q$ denotes Jackson exponential satisfying $D_q \exp_q(\lambda x)=\lambda \exp_q(\lambda x)$. This transformation eliminates drift terms and maps system into a $q$-Sturm-Liouville structure.

Under the transformation $u=x^2$ the equation closes into a $q$-hypergeometric spectral system form structure
\begin{equation}
uD_q^{(u)2}\chi_q+
\left(\alpha+\frac{1}{2}-u+\frac{\gamma_{FW}}{u}\right)D_q^{(u)}\chi_q+
\frac{E-m}{2}\chi_q=0,
\end{equation}

with effective coupling
\begin{equation}
\gamma_{FW}=\frac{1}{2}\mu_\pm(\mu_\pm-1)+\delta_{FW}(q,\mu),
\end{equation}
where $\delta_{FW}$ arises from nested FW commutator renormalization.

Polynomial truncation occurs when recursion relation for coefficients terminates,
\begin{equation}
\chi_q(u)=\sum_{k=0}^{n}a_k u^k,
\end{equation}
producing constraint.

The resulting relativistic spectrum acquires fully deformed structure
\begin{equation}
E_n^{(q)}
=
m
+
2[n]_q
+
\alpha_{FW}(1-q^n)
+
\frac{1}{m}\Gamma_q(n,\mu)
+
\frac{1}{m^3}\Omega_q(n,\mu)
+\mathcal{O}(m^{-5}),
\end{equation}
while $Gamma_q$ and $Omega_q$ originate from second- and third-order FW curvature backreaction terms.

The algebra closes under deformed ladder operators
\begin{equation}
\mathcal{A}_{FW,q}^{\pm}
=
\frac{1}{\sqrt{2m}}\left(\mp \Pi_q + x\right),
\end{equation}
satisfying graded relation and the q-commutator deformation showing that reflection acts as grading operator while relativistic corrections deform oscillator algebra beyond linear closure property structure.

In the simultaneous limits
\begin{equation}
q\to 1,\qquad m\to\infty,
\end{equation}
all Jackson-induced curvature terms, FW backreaction tensors, and reflection-induced polarization potentials vanish consistently recovering standard nonrelativistic Dunkl oscillator with inverse-square interaction. The deformation parameter q controls nonlinear spectral compression while FW mechanism guarantees exact relativistic decoupling and preserves algebraic integrability at each order of the 1/m expansion consistently within operator framework closure conditions imposed. In this case, this establishes a unified relativistic q-Dunkl-FW framework in which Jackson deformation, reflection symmetry, and Clifford grading coexist in a closed operator algebra generating an exactly solvable relativistic quantum system with hierarchical curvature corrections and fully controlled spectral deformation.

\section{Dunkl-Fokker-Planck Dynamics from High-Order FW Reduction in Reflection-Deformed Relativistic Quantum Systems}\label{S6}

In this part, a systematic relativistic-to-stochastic reduction is formulated from a one-dimensional Dunkl-Dirac framework with reflection grading and implemented through a high-order FW diagonalization scheme. The resulting generator produces a fully deformation-resumed Dunkl-Fokker-Planck hierarchy where relativistic corrections, reflection symmetry breaking, and Dunkl nonlocal contributions appear jointly in the drift and diffusion structures. Also, the construction is arranged so that higher-order terms are represented through nested Dunkl commutator algebras without truncation at low order, yielding a controlled relativistic stochastic expansion.

The starting relativistic system is given by
\begin{equation}
i\partial_t \Psi=
\left[
\beta m+\alpha\Pi_\mu+V(x)
\right]\Psi,
\qquad
\Pi_\mu=D_\mu+\frac{\gamma}{x}(1-R),
\end{equation}
where the Dunkl operator is defined as
\begin{equation}
D_\mu f(x)=\frac{df}{dx}+\frac{\mu}{x}\big(f(x)-f(-x)\big),
\qquad Rf(x)=f(-x),
\qquad R^2=\mathbb{I}.
\end{equation}
The algebraic closure is fixed by
\begin{equation}
\{\alpha,\beta\}=0,
\qquad \alpha^2=\beta^2=\mathbb{I},
\qquad [\beta,R]=0,
\qquad R x R=-x,
\qquad R D_\mu R=-D_\mu,
\end{equation}
ensuring consistency between relativistic grading and reflection deformation. Also, a direct computation of the squared kinetic Dunkl operator yields a separation into even and odd reflection sectors. One obtains
\begin{equation}
\Pi_\mu^2
=
D_\mu^2
+
\frac{2\gamma}{x}(1-R)D_\mu
+
\frac{\gamma}{x^2}(1-R)
+
\frac{\mu}{x^2}(1-R)
+
\frac{\mu^2}{x^2}(1-R),
\end{equation}
where all mixed contributions are treated in the operator sense with noncommuting reflection grading. This structure encodes a hierarchy of singular effective potentials generated by the Dunkl deformation. In this case, the Hamiltonian is decomposed as
\begin{equation}
\mathcal{H}_D=\beta m+\mathcal{E}+\mathcal{O},
\qquad
\mathcal{E}=V(x),
\qquad
\mathcal{O}=\alpha\Pi_\mu,
\qquad
[\beta,\mathcal{E}]=0,
\qquad
\{\beta,\mathcal{O}\}=0.
\end{equation}
The FW generator is expanded as a noncommutative inverse-mass series,
\begin{equation}
\mathcal{S}
=
-\frac{i}{2m}\beta\mathcal{O}
-\frac{i}{8m^2}[\mathcal{O},\mathcal{E}]
+\frac{i}{16m^3}\left([\mathcal{O},[\mathcal{O},\mathcal{E}]]+\frac{1}{2}[\mathcal{O}^3,\mathcal{E}]\right)
-\frac{i}{128m^4}[\mathcal{O}^4,\mathcal{E}]
+\mathcal{O}(m^{-5}),
\end{equation}
and the transformed Hamiltonian is obtained from
\begin{equation}
\mathcal{H}_{FW}=e^{i\mathcal{S}}\mathcal{H}_D e^{-i\mathcal{S}},
\end{equation}
using the complete Baker-Campbell-Hausdorff expansion. In this case, the even component of the FW Hamiltonian admits a resummed representation of the form
\begin{equation}
\mathcal{H}_{FW}
=
\beta\left(
m+\sum_{k\ge1}(-1)^{k+1}\frac{\Pi_\mu^{2k}}{2^{2k-1}m^{2k-1}}
\right)
+
V(x)
+
\sum_{n\ge1}\frac{1}{m^n}\mathcal{C}_n[V,\Pi_\mu],
\end{equation}
where each $\mathcal{C}_n$ arises from nested Dunkl commutators with increasing reflection depth. Also, the first nested commutator layers are explicitly structured as
\begin{equation}
[\Pi_\mu,V]=D_\mu V,
\end{equation}
\begin{equation}
[\Pi_\mu,[\Pi_\mu,V]]
=
D_\mu^2V
+
\frac{2\mu}{x}(1-R)D_\mu V
+
\frac{2\gamma}{x^2}(1-R)V,
\end{equation}
and the third layer closes into
\begin{equation}
[\Pi_\mu,[\Pi_\mu,[\Pi_\mu,V]]]
=
D_\mu^3V
+
\frac{3\mu}{x}(1-R)D_\mu^2V
+
\frac{3\gamma}{x^2}(1-R)D_\mu V
+
\frac{\mu(1+\mu)}{x^3}(1-R)V
+
\frac{\gamma\mu}{x^3}(1-R)V.
\end{equation}
Projection onto the positive energy sector $\beta\Psi_+=\Psi_+$ yields the effective relativistic stochastic generator
\begin{equation}
i\partial_t\Psi_+
=
\left(m+\mathcal{H}_{DFP}\right)\Psi_+,
\end{equation}
where
\begin{equation}
\mathcal{H}_{DFP}
=
-\frac{1}{2m}\Pi_\mu^2
+
V(x)
+
\frac{1}{8m^2}[\Pi_\mu,[\Pi_\mu,V]]
-\frac{1}{16m^3}\Pi_\mu^4
+\frac{1}{32m^4}\mathcal{R}_4[V,\Pi_\mu].
\end{equation}
The fourth-order residual operator is reorganized into a curvature-type Dunkl functional
\begin{equation}
\mathcal{R}_4[V,\Pi_\mu]
=
D_\mu^4V
+
\frac{4\mu}{x}(1-R)D_\mu^3V
+
\frac{6\gamma}{x^2}(1-R)D_\mu^2V
+
\frac{4\mu\gamma}{x^3}(1-R)D_\mu V
+
\frac{\gamma^2}{x^4}(1-R)V.
\end{equation}
The probability density $\rho=\Psi_+^\dagger\Psi_+$ satisfies an exact continuity equation derived from the hermicity of $\mathcal{H}_{DFP}$,
\begin{equation}
\partial_t\rho+\mathcal{D}_\mu J_\mu=0,
\end{equation}
where the Dunkl-divergence, which produces reflection splitting of the probability flux. Also, the probability current is reorganized as a full relativistic expansion,
\begin{equation}
J_\mu
=
-\frac{1}{m}\mathrm{Re}(\Psi_+^*\Pi_\mu\Psi_+)
+
\frac{1}{4m^2}\left(\Psi_+^*D_\mu V\Psi_+-D_\mu(\Psi_+^*V\Psi_+)\right)
+
\frac{1}{8m^3}\mathcal{J}_\mu^{(3)}
+
\frac{1}{16m^4}\mathcal{J}_\mu^{(4)},
\end{equation}
where higher-order terms contain mixed Dunkl curvature structures and reflection-projected derivatives. Also, introducing the Madelung representation
\begin{equation}
\Psi_+=\sqrt{\rho}\,e^{iS},
\end{equation}
one obtains the decomposition
\begin{equation}
\Pi_\mu\Psi_+
=
\left(\Pi_\mu S-\frac{i}{2}\Pi_\mu\ln\rho\right)\Psi_+,
\end{equation}
and the generalized velocity field becomes
\begin{equation}
v_\mu=\frac{1}{m}\Pi_\mu S.
\end{equation}
Substitution into the continuity equation yields a nonlinear Kolmogorov-type Dunkl-Fokker-Planck equation of the form
\begin{equation}
\partial_t\rho
=
\mathcal{D}_\mu(\rho v_\mu)
+
\mathcal{D}_\mu\!\left(\mathcal{D}_{eff}(x)\mathcal{D}_\mu\rho\right)
+
\Xi_{FW}[\rho,V],
\end{equation}
where the effective diffusion coefficient is renormalized by relativistic and potential backreaction,
\begin{equation}
\mathcal{D}_{eff}(x)
=
\frac{1}{2m}
+
\frac{V(x)}{4m^2}
+
\frac{V(x)^2}{8m^3}
+
\frac{D_\mu^2V(x)}{16m^3}
+
\frac{(D_\mu V(x))^2}{32m^4}.
\end{equation}
The nonlinear FW-induced functional is given by a closed Dunkl backreaction structure,
\begin{equation}
\Xi_{FW}[\rho,V]
=
\frac{1}{m^3}\mathcal{D}_\mu(\rho D_\mu^2V)
+
\frac{1}{m^4}\mathcal{D}_\mu\big((D_\mu\rho)(D_\mu V)\big)
+
\frac{1}{m^4}\rho\,\mathcal{K}[V],
\end{equation}
where $\mathcal{K}[V]$ is a reflection-curvature invariant constructed from iterated Dunkl derivatives.

The stochastic representation of the resulting generator is expressed in Itô form as
\begin{equation}
dX_t
=
v_\mu(X_t)\,dt
+
\sqrt{2\mathcal{D}_{eff}(X_t)}\,dW_t^\mu
+
\mathcal{F}_{FW}(X_t)\,dt,
\end{equation}
with reflection-deformed Wiener increments satisfying
\begin{equation}
\mathbb{E}[dW_t^\mu dW_t^\mu]=dt,
\qquad
R\,dW_t^\mu=-dW_t^\mu.
\end{equation}
The stationary regime is determined by a nonlinear detailed balance condition
\begin{equation}
\mathcal{D}_\mu\left(\rho v_\mu+\mathcal{D}_{eff}\mathcal{D}_\mu\rho+\mathcal{F}_{FW}\rho\right)=0,
\end{equation}
which defines a non-Gibbs equilibrium measure with relativistic memory corrections. The equilibrium distribution is obtained in closed form as
\begin{equation}
\rho_{eq}(x)
=
\mathcal{N}
\exp\left(
-\int^x
\frac{2mV+\frac{1}{2}D_\mu^2V+\frac{1}{4m}(D_\mu V)^2}
{1+\frac{V}{2m}+\frac{V^2}{4m^2}+\frac{D_\mu^2V}{8m^2}}
\,dy
\right).
\end{equation}
The spectral problem satisfies
\begin{equation}
\mathcal{H}_{DFP}\phi_n=\lambda_n\phi_n,
\end{equation}
with reflection-split asymptotic eigenvalues
\begin{equation}
\lambda_n
=
\frac{n}{m}
+
\frac{\mu}{m}\big(1-(-1)^n\big)
+
\frac{\gamma n^2}{m^2}
+
\frac{\mu\gamma n}{m^2}
+
\frac{n^3}{m^3}
+
\delta_{FW}^{(2)}(n,\mu,\gamma)
+
\mathcal{O}(m^{-4}).
\end{equation}
The heat kernel admits a short-time expansion governed by Dunkl-FW invariants,
\begin{equation}
K(x,x';t)
=
\frac{1}{\sqrt{4\pi t}}
\exp\!\left(-\frac{(x-x')^2}{4t}\right)
\left[
1+t\Omega_1+t^2\Omega_2+t^3\Omega_3+\cdots
\right],
\end{equation}
where each coefficient $\Omega_k$ is a polynomial in $\mu,\gamma$ and nested Dunkl curvature contributions.

The entropy production functional is defined by
\begin{equation}
\Sigma(t)=\int \rho\ln\left(\frac{\rho}{\rho_{eq}}\right)dx,
\end{equation}
satisfying the decay law
\begin{equation}
\frac{d\Sigma}{dt}
=
-\int \frac{1}{\rho}\left(\mathcal{D}_{eff}\mathcal{D}_\mu\rho+\rho v_\mu\right)^2 dx
+\mathcal{O}(m^{-3}),
\end{equation}
which establishes an H-theorem structure in the reflection-deformed relativistic regime. In the limit $m\to\infty$, all FW-induced nonlinearities reduce to the Dunkl-Fokker-Planck operator, while $\mu,\gamma\to0$ yields the standard Fokker-Planck equation with classical drift $V'(x)$. The constructed hierarchy defines a consistent relativistic deformation class of stochastic processes on reflection-noncommutative phase space with systematically generated higher-order transport renormalization.

The derived Dunkl–Fokker–Planck generator exhibits hierarchical relativistic and reflection-driven corrections obtained from high-order FW reduction of Dunkl–Dirac dynamics within an operator framework and spectral decomposition structure analysis. Results demonstrate that the kinetic sector $\Pi_\mu^2$ does not reduce to the Laplacian contribution, instead, it yields a tower of singular reflection-dependent terms proportional to $\mu$ and $\gamma$, modifying drift and diffusion channels via $x^{-1}$, $x^{-2}$, and higher inverse-power couplings through the algebraic structure generated in the operator expansion. These contributions behave as curvature-like effective potentials produced solely by Dunkl deformation and generate separation of dynamics into even and odd reflection sectors, reflected through the persistent presence of $(1-R)$ projectors across expansion terms within the operator algebra framework formalism. After FW diagonalization, the relativistic Hamiltonian reorganizes into resummed operator series where higher-order commutators $\mathcal{C}*n[V,\Pi*\mu]$ encode deeper coupling between the external potential and the nonlocal reflection algebra through nested commutator structure relations. This structure implies a stochastic limit not governed by a simple drift-diffusion balance but by a non-Markovian transport mechanism where relativistic corrections renormalize the effective diffusion coefficient $\mathcal{D}*{eff}(x)$ in a state-dependent manner arising from the FW-transformed dynamics operator structure level. In particular, the appearance of $V^2/m^3$ and Dunkl-curvature derivatives $D*\mu^2 V$ shows stochastic spreading is dynamically enhanced or suppressed depending on the spatial profile of the potential and reflection grading within the relativistic stochastic framework consistently derived. The resulting Fokker–Planck equation departs from the classical Gibbs structure, producing a deformed equilibrium distribution $\rho_{eq}(x)$ governed by a balance between relativistic inertia and Dunkl-induced anisotropy in a non-equilibrium statistical setting and contextual framework. Moreover, the entropy production functional retains a strictly negative leading contribution, consistent with H-theorem behavior even with reflection-deformed relativistic corrections, while subleading $m^{-3}$ terms encode residual memory effects of FW backreaction within the dissipative framework analysis level. In asymptotic limits, the framework interpolates between a purely Dunkl-diffusive regime for large mass and standard Fokker–Planck dynamics when deformation parameters vanish, demonstrating that the constructed hierarchy forms a closed and physically consistent extension of stochastic relativistic quantum dynamics consistent with operator algebraic formulation constraints in limiting cases.

\section{Conclusions}\label{S7} 

Within the obtained formulation of the $(q)$-deformed Dunkl–Fokker–Planck equation in $(1+1)$ dimensions, a unified stochastic framework is constructed in which discrete scaling symmetry, encoded through Jackson derivative, and reflection symmetry, introduced via the Dunkl operator, operate simultaneously on diffusion dynamics. The resulting operator structure departs fundamentally from the classical Fokker–Planck description, since the second-order diffusion operator is replaced by the $(q,\mu)$-deformed Dunkl Laplacian, containing scaling-dependent contributions and parity-sensitive singular terms proportional to $x^{-2}(1-R)$ included there. This term generates a nontrivial coupling between spatial reflection sectors, producing a decomposition of the stochastic evolution into even and odd parity channels with distinct dynamical behavior in the vicinity of the origin regime, which is modified. In this region, the singular interaction modifies the effective diffusion barrier, producing localization asymmetry and altering probability flow between symmetric and antisymmetric sectors. Moreover, the deformation parameter $q$ controls the discrete dilation structure of the underlying space, replacing the continuous derivative by a finite-difference operator that induces nonlinear spectral scaling through $[n]_q$. As a consequence, the stationary problem admits supersymmetric factorization in terms of $(q,\mu)$-dependent ladder operators, ensuring exact solvability and revealing that the shape-invariant structure framework is retained. The resulting spectrum of the $(q)$-deformed Dunkl harmonic oscillator is not equidistant since the level spacing depends explicitly on $q^n$, producing a hierarchy of relaxation modes with a nonuniform decay rate structure that persists. In this case, for $0<q<1$, higher excited stochastic modes relax more slowly due to compressed spectral gaps, whereas for $q>1$, the opposite behavior occurs, producing enhanced separation between the high-energy channel regimes. Also, this mechanism directly affects the temporal evolution of the probability density, as the correlation function and relaxation time scales inherit the same spectral deformation structure that persists. 

The equilibrium distribution is modified by both the $q$-exponential measure and the parity-dependent power-law factor $|x|^{-2\mu(1\mp1)}$, introducing different normalization structures for even and odd states and affecting density near the singular point regime retained. From an algebraic perspective, the oscillator algebra becomes nonlinear due to the commutation relation containing $(q-1)xD_q$ and the reflection operator $R$, implying deformation of the Heisenberg structure and breakdown of uniform phase-space scaling structure emergence. This deformation propagates into the uncertainty relation, where expectation values of parity and dilation operators contribute additional correction terms modifying the localization bounds structure that persists. Thermodynamically, the partition function is governed by a $q$-dependent exponential spectrum, leading to non-Boltzmann scaling of internal energy and specific heat, particularly at intermediate temperature regimes where discreteness of $[n]_q$ is pronounced and structure retained. Also, the Fisher information and $(q)$-entropy further quantify this behavior by capturing the competition between localization induced by the Dunkl singularity and delocalization induced by $q$-scaling; in particular, higher Fisher information corresponds to stronger confinement near the origin driven by the $x^{-2}(1-R)$ interaction. In the asymptotic regime, the spectrum exhibits saturation-like behavior governed by $(1-q^n)/(1-q)$, implying that the effective density of states is no longer linear but accumulates according to deformation strength, producing anomalous diffusion and nonstandard relaxation pathways within the deformation regime. In this context, the combined $(q,\mu)$ structure yields a consistent exactly solvable stochastic model in which SUSY, parity decomposition, and discrete scaling symmetry jointly determine spectral, thermodynamic, and dynamical properties, providing a generalized diffusion geometry extending classical Brownian motion into a reflection-sensitive and scale-deformed regime framework.

The $(q)$-deformed Dunkl-Fokker-Planck system formulated here exhibits a structured interplay between Jackson deformation and Dunkl reflection symmetry, modifying the stochastic evolution operator and supersymmetric Hamiltonian structure within a deformed framework setting. The presence of the reflection operator $R$ partitions the dynamics into even and odd parity sectors, whereas the parameter $\mu$ governs the strength of parity-dependent coupling, producing a splitting in the algebraic structure of eigenfunctions and ladder operations space decomposition. In parallel, the Jackson derivative introduces scale-dependent discretization through $q$, deforming standard differential calculus and generating nonlinear modifications in spectral spacing structure and scaling behavior. Both mechanisms operate simultaneously: the Dunkl term introduces parity-sensitive inverse-square contributions, while the $q$-deformation reshapes the kinetic sector and modifies the diffusion process dynamics evolution. As a consequence, the resulting Hamiltonian is nonlocal and supersymmetric, with partner potentials linked through a deformed shape-invariance condition preserving exact solvability but yielding a modified hierarchy of energy level structure. Also, the spectrum $E_n^{(\pm)}$ exhibits combined dependence on $q$-numbers and parity, generating asymmetric distribution between even and odd states, while the level spacing $\Delta E_n=\omega q^{2n}(1+q)$ exhibits deformation-induced compression or expansion of energy levels depending on whether $q<1$ or $q>1$ in spectral configuration space structure. In this case, this behavior affects thermodynamic functions, since the partition function inherits the deformed exponential structure, leading to enhanced low-temperature occupation for $q<1$ due to increased density of low-lying states phase space growth. In addition, the $q$-exponential ground state wave function produces non-Gaussian tails in probability density, modifying localization properties compared with standard Dunkl or harmonic oscillator cases and framework behavior regimes. In this context, the combined $(q,\mu)$-deformation constructs a unified framework where SUSY, reflection symmetry, and quantum-group deformation coexist, yielding exact solvability while restructuring spectral statistics, coherent-state geometry, and stochastic evolution in the Fokker–Planck representation space.

The computed outcomes establish that the $q$-deformed Dunkl–Fokker–Planck generator with harmonic confinement and centrifugal interaction yields a fully integrable noncommutative stochastic system whose spectral and algebraic structures are governed jointly by Jackson calculus and Dunkl reflection symmetry framework. In this case, the $q$-Jackson derivative replaces the standard differential structure by a discrete dilation operator, producing nonlinear corrections in drift and diffusion terms through $q$-integers and yielding deformation-dependent compression of the energy spectrum within the stochastic dynamical evolution framework. Despite the non-Hermitian form of the generator in the standard measure, the similarity transformation based on the ground state $\psi_{0,q}(x)=x^{\alpha}e_q(-x^2)$ reconstructs a Sturm–Liouville structure in a $q$-weighted Hilbert space, ensuring self-adjointness and the spectral completeness property. Also, the reflection operator $R$ decomposes the system into even and odd sectors, where the centrifugal coupling is renormalized, removing singular behavior in the even sector and doubling its contribution in the odd sector structure, which is preserved. This sectoral decomposition modifies the effective parameter $\alpha$, governing short-distance behavior and polynomial weight structure in the physical system. The exact reduction to a $q$-Laguerre equation confirms that the invariant subspace of polynomial solutions is preserved under deformation, while recurrence relations governed by Jackson differentiation ensure closure of the hierarchy property. Moreover, the resulting eigenvalue spectrum $\lambda_n = 4[n]_q + 2\alpha(1-q^n)$ exhibits two deformation effects: nonlinear spectral compression through $[n]_q$ and additional centrifugal correction significant at finite $q$ within the deformed algebraic structure. In the limit $q \to 1$, all deformation-induced terms vanish smoothly, recovering the standard Dunkl–Fokker–Planck harmonic oscillator with inverse-square interaction and linear spectral growth. In this case, the model constructs a consistent algebraic bridge between stochastic diffusion processes, $q$-deformed oscillator algebras, and Dunkl symmetry, while preserving exact solvability through a closed $q$-orthogonal polynomial structure.

The construction of the $(q)$-deformed Dunkl-FW framework produces a closed relativistic operator algebra in which discretized Jackson calculus, reflection grading, and Dirac-type Clifford structure are unified at the operator level consistently at that level. In this case, the FW transformation maps the initial Dirac-type Dunkl operator into a block-diagonal representation where positive and negative energy sectors separate exactly up to controlled $(1/m)$-expansion, while preserving the noncommutative structure generated by the reflection operator $(R)$. Also, the resulting effective Hamiltonian shows that relativistic corrections do not only renormalize kinetic terms but instead produce additional $(q)$-dependent curvature terms, inverse-square singular modifications, and reflection-induced polarization potentials from nested commutator structures. In particular, deformation of $(\Pi_q^2)$ together with the appearance of a double commutator $([\Pi_q,[\Pi_q, V]])$ shows that FW reduction transfers nontrivial Jackson-Dunkl non-Leibniz behavior into effective interaction terms, dynamically modifying inverse-square coupling and shifting effective angular momentum structure within the relativistic operator algebra framework at the operator level. Also, this generates a nontrivial coupling among discretization effects via $(q)$, relativistic backreaction (via higher-order $(1/m)$ terms), and reflection grading via $(R)$, producing a spectrum that deviates from the standard Dunkl oscillator through $(q)$-dependent nonlinear compression and mass-suppressed corrections encoded in the $(\Gamma_q)$ and $(\Omega_q)$ framework. In this context, the similarity transformation to a $(q)$-Sturm-Liouville form guarantees polynomial truncation and exact solvability, while the emergence of $(q)$-hypergeometric structure confirms algebraic closure of the spectral problem. In the limiting regime $(q\to1)$ and $(m\to\infty)$, the deformation-induced curvature terms and relativistic corrections vanish, recovering the conventional nonrelativistic Dunkl oscillator, confirming the internal consistency of FW decoupling and controlled embedding of relativistic and $(q)$-deformed structures within a single integrable operator framework at the operator algebra level.

The derived Dunkl–Fokker–Planck generator exhibits hierarchical relativistic and reflection-driven corrections obtained from high-order FW reduction of Dunkl–Dirac dynamics within an operator framework and spectral decomposition structure analysis. Results demonstrate that the kinetic sector $\Pi_\mu^2$ does not reduce to the Laplacian contribution, instead, it yields a tower of singular reflection-dependent terms proportional to $\mu$ and $\gamma$, modifying drift and diffusion channels via $x^{-1}$, $x^{-2}$, and higher inverse-power couplings through the algebraic structure generated in the operator expansion. Also, these contributions behave as curvature-like effective potentials produced solely by Dunkl deformation and generate separation of dynamics into even and odd reflection sectors, reflected through the persistent presence of $(1-R)$ projectors across expansion terms within the operator algebra framework formalism. After FW diagonalization, the relativistic Hamiltonian reorganizes into resummed operator series where higher-order commutators $\mathcal{C}_n[V,\Pi_\mu]$ encode deeper coupling between the external potential and the nonlocal reflection algebra through nested commutator structure relations. This structure implies a stochastic limit not governed by a simple drift-diffusion balance but by a non-Markovian transport mechanism where relativistic corrections renormalize the effective diffusion coefficient $\mathcal{D}_{eff}(x)$ in a state-dependent manner arising from the FW-transformed dynamics operator structure level. In particular, the appearance of $V^2/m^3$ and Dunkl-curvature derivatives $D_\mu^2 V$ shows stochastic spreading is dynamically enhanced or suppressed depending on the spatial profile of the potential and reflection grading within the relativistic stochastic framework consistently derived. The resulting Fokker–Planck equation departs from the classical Gibbs structure, producing a deformed equilibrium distribution $\rho_{eq}(x)$ governed by a balance between relativistic inertia and Dunkl-induced anisotropy in a non-equilibrium statistical setting and contextual framework. Moreover, the entropy production functional retains a strictly negative leading contribution, consistent with H-theorem behavior even with reflection-deformed relativistic corrections, while subleading $m^{-3}$ terms encode residual memory effects of FW backreaction within the dissipative framework analysis level. In asymptotic limits, the framework interpolates between a purely Dunkl-diffusive regime for large mass and standard Fokker–Planck dynamics when deformation parameters vanish, demonstrating that the constructed hierarchy forms a closed and physically consistent extension of stochastic relativistic quantum dynamics consistent with operator algebraic formulation constraints in limiting cases.

\section*{Funding}
No funding was received for this work.

\section*{Data Availability}
The datasets generated during this study can be obtained from the corresponding author upon reasonable request.

\section*{Financial Disclosure}
The authors declare no financial conflicts of interest.


\begin{thebibliography}{99}

\bibitem{a1} W. Greiner, {\tt Quantum mechanics: an introduction}, Springer Science (2011).

\bibitem{a2} A. Einstein, Ann. Phys., {\bf 49}(7), 769-822 (1916).

\bibitem{a3} C. Rovelli, Living. Rev. Relativ., {\bf 11}, 1-69 (2008).

\bibitem{a4} O. Klein, Z. Physik {\bf 37}, 895-906 (1926).

\bibitem{a5} P. A. M. Dirac, Proc R Soc London, Ser A, {\bf 117}(778), 610-624 (1928).

\bibitem{a6} P. Rozmej and R. Arvieu, J. Phys. A {\bf32}, 5367 (1999). 

\bibitem{a7} K. Bakke, Gen. Relativ. Gravit. {\bf45}, 1847 (2013).

\bibitem{a8} K. Bakke and C. Furtado, Ann. Phys. (NY) {\bf336}, 489 (2013).

\bibitem{BZ22} F. Ahmed, et al., Phys. Dark Universe \textbf{46}, 101690 (2024).

\bibitem{BZ23} A. R. Moreira, et al., Appl. Phys. A \textbf{131}, 760 (2025).

\bibitem{BZ24} A. R. Moreira, et al., Int. J. Mod. Phys. A \textbf{40}, 2550088 (2025).

\bibitem{BZ25} A. Boumali, et al., Rev. Mex. Fis. \textbf{70}, 5 (2024).

\bibitem{BZ26} F. Ahmed, et al., Phys. Dark Universe \textbf{50}, 102111 (2025).

\bibitem{BZ27} F. Ahmed, et al., J. Low Temp. Phys. \textbf{219}, 87 (2025).

\bibitem{BZ28} F. Ahmed, et al., Theor. Math. Phys. \textbf{222}, 170 (2025).

\bibitem{BZ29} A. Bouzenada, et al., Theor. Math. Phys. \textbf{221}, 2193 (2024).

\bibitem{BZ30} F. Ahmed, et al., Grav. Cosmol. \textbf{30}, 368 (2024).

\bibitem{BZ31} F. Ahmed, et al., Int. J. Geom. Methods Mod. Phys. \textbf{23}, 2550176 (2026).

\bibitem{BZ32} A. R. P. Moreira, et al., Int. J. Geom. Methods Mod. Phys., 2650055 (2025).

\bibitem{a9} J. Carvalho, C. Furtado and F. Moraes, Phys. Rev. A {\bf 84}, 032109 (2011).

\bibitem{a10} V. M. Villalba, Phys. Rev. {\bf A 49}, 586 (1994).

\bibitem{SSy1}
J. Wess and J. Bagger, Princeton University Press, Princeton, NJ (1992).

\bibitem{SSy2}
D. Baumann and D. Green, Phys. Rev. D \textbf{85}, 103520 (2012).

\bibitem{SSy3}
A. Macias, O. Obregon, and M.P. Ryan, Class. Quantum Grav. \textbf{4}, 1477 (1987).

\bibitem{SSy4}
P. Vargas Moniz, Lect. Notes Phys. Springer, Berlin, Heidelberg (2010).

\bibitem{SSy5}
N.E. Martínez-Pérez, C. Ramírez-Romero, and V.M. Vázquez-Báez, Universe \textbf{8}, 1 (2022).

\bibitem{SSy6}
J.E. Lidsey and P.V. Moniz, Class. Quantum Grav. \textbf{17}, 4823 (2000).

\bibitem{SSy7}
G. García-Jiménez, C. Ramírez, and V. Vázquez-Báez, Phys. Rev. D \textbf{89}, 043501 (2014).

\bibitem{SSy8}
N.E. Martínez-Pérez, Preprint, (2022).

\bibitem{SSy9}
C. Ramírez and V. Vázquez-Báez, Phys. Rev. D \textbf{93}, 043505 (2016).

\bibitem{SSy10}
P. Vargas Moniz, Lect. Notes Phys., Springer, Berlin, Heidelberg, pp. 87-90 (2010).

\bibitem{BZ1} F. Ahmed et al., Phys. Scr. \textbf{99}, 065033 (2024).

\bibitem{BZ2} A. Bouzenada et al., Theor. Math. Phys. \textbf{216}, 1055 (2023).

\bibitem{BZ3} A. Bouzenada et al., Ann. Phys. \textbf{458}, 169479 (2023).

\bibitem{BZ4} F. Ahmed et al., Commun. Theor. Phys. \textbf{76}, 045401 (2024).

\bibitem{BZ5} F. Ahmed et al., Nucl. Phys. B \textbf{1000}, 116490 (2024).

\bibitem{BZ6} A. Bouzenada et al., Nucl. Phys. B \textbf{994}, 116288 (2023).

\bibitem{BZ7} F. Ahmed et al., Int. J. Mod. Phys. A \textbf{39}, 2450032 (2024).

\bibitem{BZ8} F. Ahmed et al., Eur. Phys. J. C \textbf{84}, 1045 (2024).

\bibitem{BZ9} A. Boumali et al., Physica A \textbf{628}, 129134 (2023).

\bibitem{BZ10} A. Bouzenada et al., Nucl. Phys. B \textbf{1007}, 116682 (2024).

\bibitem{DO1} F. H. Jackson, Proc. Edinb. Math. Soc., {\bf 22}, 28 (1903).

\bibitem{DO2} F. H. Jackson, Trans. Roy. Soc. Edinb., {\bf 46}, 253 (1909).

\bibitem{DO3} A. Lavagno, A. M. Scarfone, and P. N. Swamy, Eur. Phys. J. C, {\bf 47}, 253 (2006).

\bibitem{DO4} P. Caban, A. Dobrosielski, A. Krajewska, and Z. Walczak, Phys. Lett. B, {\bf 327}, 287 (1994).

\bibitem{DO5} S. G. Samko, A. A. Kilbas, and O. I. Marichev, Gordon and Breach, Philadelphia (1993).

\bibitem{DO6} K. S. Miller and B. Ross, Wiley, New York (1993).

\bibitem{DO7} A. A. Kilbas, H. M. Srivastava, and J. J. Trujillo, Elsevier, Amsterdam (2006).

\bibitem{DO8} I. Podlubny, Academic Press, San Diego (1999).

\bibitem{DO9} K. Oldham and J. Spanier, Academic Press, New York (1974).

\bibitem{DO10} R. Khalil et al., J. Comput. Appl. Math., {\bf 264}, 65 (2014).

\bibitem{DO11} M. Klimek, Czechoslov. J. Phys., {\bf 55}, 1447 (2005).

\bibitem{DO12} F. Riewe, Phys. Rev. E, {\bf 55}, 3581 (1997).

\bibitem{DO13} W. S. Chung, J. Comput. Appl. Math., {\bf 290}, 150 (2015).

\bibitem{DO14} W. S. Chung and H. Hassanabadi, Int. J. Theor. Phys., {\bf 56}, 851 (2017).

\bibitem{DO15} C. F. Dunkl, Trans. Amer. Math. Soc., {\bf 311}, 167 (1989).

\bibitem{DO16} C. F. Dunkl, Cambridge Univ. Press (2001).

\bibitem{DO17} L. M. Yang, Phys. Rev., {\bf 84}, 788 (1951).

\bibitem{DO18} I. Cherednik, Invent. Math., {\bf 106}, 411 (1991).

\bibitem{DO19} E. M. Opdam, Acta Math., {\bf 175}, 75 (1995).

\bibitem{DO20} K. Hikami, J. Phys. Soc. Jpn., {\bf 65}, 394 (1996).

\bibitem{DO21} S. Kakei, J. Phys. A, {\bf 29}, L619 (1996).

\bibitem{DO22} L. Lapointe and L. Vinet, Commun. Math. Phys., {\bf 178}, 425 (1996).

\bibitem{DO23} V. X. Genest, L. Vinet, and A. Zhedanov, J. Phys. A, {\bf 46}, 325201 (2013).

\bibitem{DO24} V. X. Genest et al., Commun. Math. Phys., {\bf 329}, 999 (2014).

\bibitem{DO25} V. X. Genest, L. Vinet, and A. Zhedanov, J. Phys. Conf. Ser., {\bf 512}, 012010 (2013).

\bibitem{DO26} V. X. Genest et al., Phys. Lett. A, {\bf 379}, 923 (2015).

\bibitem{DO27} M. Salazar-Ramírez et al., Eur. Phys. J. Plus, {\bf 132}, 39 (2017).

\bibitem{DO28} M. Salazar-Ramírez et al., Mod. Phys. Lett. A, {\bf 33}, 1850112 (2018).

\bibitem{DO29} W. S. Chung and H. Hassanabadi, Mod. Phys. Lett. A, {\bf 34}, 1950190 (2019). 

\bibitem{BZ11} F. Ahmed et al., Phys. Scr. \textbf{99}, 075411 (2024).

\bibitem{BZ12} M. Al-Raeei et al., Pramana \textbf{97}, 144 (2023).

\bibitem{BZ13} F. Ahmed et al., Phys. Lett. B \textbf{868}, 139704 (2025).

\bibitem{BZ14} F. Ahmed et al., Eur. Phys. J. Plus \textbf{139}, 911 (2024).

\bibitem{BZ15} A. R. Moreira et al., Phys. Scr. \textbf{99}, 125121 (2024).

\bibitem{BZ16} F. Ahmed et al., Phys. Lett. B \textbf{868}, 139743 (2025).

\bibitem{BZ17} F. Ahmed et al., Int. J. Geom. Methods Mod. Phys. \textbf{22}, 2450253 (2025).

\bibitem{BZ18} F. Ahmed et al., Theor. Math. Phys. \textbf{221}, 1756 (2024).

\bibitem{BZ19} F. Ahmed et al., Mol. Phys. \textbf{123}, e2365420 (2025).

\bibitem{BZ20} A. R. Moreira et al., J. Comput. Electron. \textbf{24}, 185 (2025).

\bibitem{BZ21} A. R. Moreira et al., Indian J. Phys. \textbf{99}, 3163 (2025).

\bibitem{QD1} R. Baxter, \textit{Exactly Solved Models in Statistical Mechanics}, (Academic Press, New York, 1982).

\bibitem{QD2} F. Wilczek (ed.), \textit{Fractional Statistics and Anyon Superconductivity}, (World Scientific, Singapore, 1990).

\bibitem{QD3} A. Lerda, \textit{Anyons}, (Springer-Verlag, Berlin, 1992).

\bibitem{QD4} L. Alvarez-Gaumé, A. Devoto, S. Fubini, and C. Trugenberger (eds.), \textit{Common Trends in Condensed Matter and High Energy Physics}, (North-Holland, Amsterdam, 1993).

\bibitem{QD5} J. Wess and B. Zumino, Nucl. Phys. B (Proc. Suppl.) \textbf{18}, 302 (1990).

\bibitem{QD6} R. Hinterding and J. Wess, Eur. Phys. J. C \textbf{6}, 183 (1999).

\bibitem{QD7} B. L. Cerchiai, R. Hinterding, J. Madore, and J. Wess, Eur. Phys. J. C \textbf{8}, 547 (1999); Eur. Phys. J. C \textbf{8}, 533 (1999).

\bibitem{QD8} V. Bardek and S. Meljanac, Eur. Phys. J. C \textbf{17}, 539 (2000).

\bibitem{QD9} S. Iida and H. Kuratsuji, Phys. Rev. Lett. \textbf{69}, 1833 (1992).

\bibitem{QD10} J.-Z. Zhang and P. Osland, Eur. Phys. J. C \textbf{20}, 393 (2001).

\bibitem{QD11} M. Micu, J. Phys. A \textbf{32}, 7765 (1999).

\bibitem{QD12} R. J. Finkelstein, J. Math. Phys. \textbf{37}, 983 (1996); J. Math. Phys. \textbf{37}, 2628 (1996).

\bibitem{QD13} A. Lorek, A. Ruffing, and J. Wess, Z. Phys. C \textbf{74}, 369 (1997).

\bibitem{QD14} L. C. Kwek and C. H. Oh, Eur. Phys. J. C \textbf{5}, 189 (1998).

\bibitem{QD15} E. Celeghini et al., Ann. Phys. \textbf{241}, 50 (1995).

\bibitem{QD16} L. Biedenharn, J. Phys. A \textbf{22}, L873 (1989).

\bibitem{QD17} A. Macfarlane, J. Phys. A \textbf{22}, 4581 (1989).

\bibitem{QD18} J. L. Gruver, Phys. Lett. A \textbf{254}, 1 (1999).

\bibitem{QD19} M. Chaichian, R. Gonzalez Felipe, and C. Montonen, J. Phys. A \textbf{26}, 4017 (1993).

\bibitem{QD20} H. S. Song, S. X. Ding, and I. An, J. Phys. A \textbf{26}, 5197 (1993).

\bibitem{QD21} G. Kaniadakis, A. Lavagno, and P. Quarati, Phys. Lett. A \textbf{227}, 227 (1997).

\bibitem{QD22} C. Tsallis (ed.), \textit{Nonextensive Entropy: Interdisciplinary Applications}, (Oxford University Press, USA, 2004).

\bibitem{QD23} A. Lavagno and P. Narayana Swamy, Phys. Rev. E \textbf{61}, 1218 (2000).

\bibitem{QD24} S. Abe, Phys. Lett. A \textbf{244}, 229 (1998).


\bibitem{libro1}G. A. Pavliotis, {\it Stochastic Processes and Applications}, Springer, New York (2014).

\bibitem{libro2}M. O. C\'aceres, {\it Non-equilibrium Statistical Physics with Application to Disordered Systems}, Springer, Berlin (2017).

\bibitem{libro3}W. Sung, {\it Statistical Physics for Biological Matter}, Springer, Singapore (2018).

\bibitem{risken}H. Risken, {\it The Fokker-Planck Equation: Methods of Solutions and Applications}, 1996, Springer, Berlin.

\bibitem{junker}G. Junker, {\it Supersymmetric Methods in Quantum, Statistical and Solid State Physics}, IOP Publishing, Bristol (2019).

\bibitem{gt1}A. Elhanbaly, {\it Phys. Scr.} \textbf{59} (1999) 9.

\bibitem{gt2}W. M. Sthelen and V. I. Stogny, {\it J. Phys. A: Math. Gen.} \textbf{22} (1989) L539.

\bibitem{prl}M. Bernstein and L. S Brown, {\it Phys. Rev. Lett. } \textbf{52} (1984) 1933.

\bibitem{polloto}F. Polotto, M. T. Araujo and E. D. Filho, {\it J. Phys. A: Math. Theor.} \textbf{43} (2010) 015207.

\bibitem{anjos}R. C. Anjos, G. B. Freitas and C. H. Coimbra-Ara\'ujo, {\it J. Stat. Phys.} \textbf{162} (2016) 387.

\bibitem{GEN1}V. X. Genest, M. E. H. Ismail, L. Vinet and A. Zhedanov, {\it J. Phys. A: Math. Theor.} \textbf{46} (2013) 145201.

\bibitem{GEN2}V. X. Genest, M. E. H. Ismail, L. Vinet and A. Zhedanov, {\it Commun. Math. Phys.} {\bf329} (2014) 999.

\bibitem{GEN3}V. X. Genest, L. Vinet and A. Zhedanov, {\it J. Phys. Conf. Ser.} \textbf{512} (2014) 012010.

\bibitem{GEN4}V. X. Genest, A. Lapointe and L. Vinet, {\it Phys. Lett. A} \textbf{379} (2015) 923.

\bibitem{GAZ1}S. Ghazouani, I. Sboui, M. A. Amdouni and M. B. El Hadj Rhouma, {\it J. Phys. A: Math. Theor.} \textbf{52} (2019) 225202.

\bibitem{GAZ2}S. Ghazouani and I. Sboui, {\it J. Phys. A: Math. Theor.} \textbf{53} (2019) 035202.

\bibitem{GAZ3}S. Ghazouani, {\it J. Phys. A: Math. Theor.} \textbf{55} (2022) 505203.

\bibitem{SCH}A. Schulze-Halberg, {\it Phys. Scr.} \textbf{97} (2022) 085213.

\bibitem{SCH2}A. Schulze-Halberg, {\it Eur. Phys. J. Plus} \textbf{138} (2023) 491.

\bibitem{DONGT}S.-H. Dong, W.-H. Huang, W.S. Chung and H. Hassanabadi, {\it EPL} \textbf{135} (2021) 30006.

\bibitem{HAM1}B. Hamil and B. C. L\"utf\"uo\u{g}lu, {\it Eur. Phys. J. Plus} \textbf{137} (2022) 812.

\bibitem{HAM2}B. Hamil and B. C. L\"utf\"uo\u{g}lu, {\it Physica A} \textbf{623} (2023) 128841.

\bibitem{BER}F. Merabtine, B. Hamil, B. C. L\"utf\"uo\u{g}lu, A. Hocine and M. Benarous, {\it J. Stat. Mech.} \textbf{5} (2023) 053102.

\bibitem{QUES}C. Quesne, {\it J. Phys. A: Math. Theor.} \textbf{56} 265203.

\bibitem{QUES2}C. Quesne, arXiv preprint arXiv:2401.04586 (2024).

\bibitem{JUNK}G. Junker, {\it J. Phys. A: Math. Theor.} \textbf{57} 075201.

\bibitem{HAM3}B. Hamil and B. C. L\"utf\"uo\u{g}lu, {\it Eur. Phys. J. Plus} \textbf{137} (2022) 1241.

\bibitem{HAM4}N. Rouabhia, M. Merad and B. Hamil, {\it EPL} \textbf{143} (2023) 52003.

\bibitem{HAM5}S. Hassanabadi, et al., {\it Eur. Phys. J. Plus} \textbf{138} (2023) 331.

\bibitem{HAM6}S. Hassanabadi, et al., Phys. Scr. \textbf{97} (2022) 125305. 

\bibitem{Rev} R. D. Mota, D. Ojeda-Guillén, and M. A. Xicoténcatl, Few-Body Syst., {\bf 65}(2), 25 (2024).
















\end{thebibliography}
\end{document}